\documentclass[11pt, a4paper]{article}
\usepackage[utf8]{inputenc}
\usepackage{amsmath}
\usepackage{centernot}
\usepackage{graphicx,color}
\usepackage{authblk}
\usepackage[natbibapa]{apacite}
\usepackage[margin=1.2in]{geometry}

\newcommand{\ER}{Erd\"{o}s-R\'enyi }

\providecommand{\keywords}[1]
{
  \small
  \textbf{\textit{Keywords ---}} #1
}

\begin{document}

\title{Network Sensitivity of Systemic Risk}

\author[1]{Amanah Ramadiah}
\author[2]{Domenico Di Gangi}
\author[ ]{D. Ruggiero Lo Sardo}
\author[3]{Valentina Macchiati}
\author[ ]{Tuan Pham Minh}
\author[4]{Francesco Pinotti}
\author[2]{Mateusz Wilinski}
\author[1]{Paolo Barucca}
\author[5,6,7]{Giulio Cimini}
\affil[1]{Department of Computer Science, University College London, London WC1E6BT, United Kingdom}
\affil[2]{Scuola Normale Superiore, 56126 Pisa, Italy}
\affil[3]{Department of Physics, University of Turin, 10125 Torino, Italy}
\affil[4]{INSERM, Sorbonne Université, Institut Pierre Louis d’Épidémiologie et de Santé Publique (IPLESP), 75012 Paris, France}
\affil[5]{Department of Physics, University of Rome Tor Vergata, 00133 Rome, Italy}
\affil[6]{IMT School for Advanced Studies, 55100 Lucca, Italy}
\affil[7]{Institute for Complex Systems (CNR), 00185 Rome, Italy}

\maketitle

\begin{abstract}
A growing body of studies on systemic risk in financial markets has emphasized the key importance of taking into consideration the complex interconnections among financial institutions. 
Much effort has been put in modeling the contagion dynamics of financial shocks, and to assess the resilience of specific financial markets---either using real network data, reconstruction techniques or simple toy networks. 
Here we address the more general problem of how shock propagation dynamics depends on the topological details of the underlying network. 
To this end we consider different realistic network topologies, all consistent with balance sheets information obtained from real data on financial institutions. 
In particular, we consider networks of varying density and with different block structures, and diversify as well in the details of the shock propagation dynamics. 
We confirm that the systemic risk properties of a financial network are extremely sensitive to its network features. 
Our results can aid in the design of regulatory policies to improve the robustness of financial markets.
\end{abstract}


\hspace{0pt}

\textbf{\textit{Key messages}}
\begin{itemize}
   \item We generalise a reconstruction method for financial networks, consistent with balance sheets information
   \item We study the interplay between the network structure and the dynamics of shocks propagation
   \item We show that systemic risk is extremely sensitive to the features of the financial network
   \item Our results can help design policies to improve the robustness of financial markets
\end{itemize}

\section{Introduction}

The crises that hit the financial world in the last two decades led scientists and regulators to rethink, with a systemic perspective, the approach used to assess market risk with an increased interest in the entangled structure of financial relationships \citep{allen2007systemic,allen2014transmission,acemoglu2015systemic}, its role in the potential propagation of risk \citep{gai2010contagion} and its consequences to risk management and macroprudential regulation 
\citep{haldane2011systemic,battiston2016complexity}.
A common denominator that emerged from the many empirical works on systemic risk has been the importance of considering the role of the structure of financial dependencies 
\citep{boss2004network,nier2007network,cocco2009lending,georg2013effect} and how centrality measures in financial networks can be crucial to identify systemically important financial institutions \citep{battiston2012debtrank}.
Whilst the evidence of the role of interconnections and the need for their direct measurement or reconstruction \citep{anand2017missing} has grown, at the same time also the research on designing novel and more realistic systemic risk mechanisms greatly developed in recent years.
Starting from the seminal works on clearing mechanisms \citep{eisenberg2001systemic} and default contagion \citep{furfine2003interbank}, now a growing number of extensions have been introduced \citep{amini2016resilience,acharya2017measuring,benoit2017where,caccioli2018network}, and we now have methods ranging from the seminal default contagion approaches to distress contagion, such as DebtRank centrality \citep{battiston2012debtrank} and network valuation \citep{barucca2016network}.
These kinds of network methodologies are nowadays implemented in stress tests and stability analysis performed by central banks \citep{bardoscia2019forward, covi2019economic}.
Therefore, the current scientific challenge is no longer to generically quantify systemic risk, but to be able to understand in more detail the interplay between systemic risk measures and network structures, as different contagion mechanisms may yield different risks and vulnerabilities depending on the underlying network structure. 
Only with this increased level of understanding, specific regulatory solutions to improve the structure of the system and reduce risk can become reliable.
To this end, it is essential to understand which features of a financial network make it more or less resilient to systemic risk.
One of the first contributions in this direction is the work by \cite{gai2010contagion}, who showed that random \ER networks are ``robust-yet-fragile'': the probability of contagion is maximal for intermediate network densities, whereas, the amount of systemic losses monotonically increases with the network connectivity.
\cite{mastromatteo2012reconstruction} further showed that, under the Furfine dynamics, sparse \ER networks in general lead to more defaults than very dense networks. 
\cite{roukny2013default} noted that no single topology can always lead to lowest risk levels (in particular, scale-free networks can be both more robust and more fragile than \ER architectures).
\cite{leon2014rethinking} argued that modular scale-free architectures can favor robustness, whereas, \cite{montagna2017contagion} pointed out that the dependence of systemic risk on the density changes if shocks are correlated.
\cite{hurd2017framework} observed that, under the Gai \& Kapadia dynamics, degree assortativity can strongly affect the course of contagion cascades \citep{hurd2017framework}.
For what concerns the DebtRank dynamics, \cite{bardoscia2017pathways} showed that the stability of the system decreases monotonically with the link density due to the presence of cycles, whereas \cite{krause2019controlling} recently pointed out that degree assortativity correlates well with the level of systemic risk. 

In this work, for the first time, we generalise one of the most consolidated methods for reconstructing realistic financial networks \cite{anand2017missing} for the case of weighted heterogeneous networks with block structure, both core-periphery and modular one.
Furthermore, we introduce a robust and efficient network sensitivity methodology which explores a range of weighted financial networks with varying density, from extremely sparse low-density networks to complete ones, and applies two paramount models of both default contagion and distress contagion, displaying the significant differences in relative losses that can arise from different network structures, shocks applied, and contagion mechanisms.

\section{Methodology}

In this section we explain the two-steps procedure we use in our framework.
In a nutshell, we firstly generate a reconstructed financial network with some key characteristics, and then run a shock propagation dynamics, Furfine and DebtRank algorithms, over it in order to assess its level of systemic risk.

\subsection{Data}

We base our reconstruction of financial (specifically, interbank) networks on Bankscope data \citep{battiston2016leveraging} containing the balance sheet of the 100 largest European banks. 
In particular, we have information about the interbank assets $A_i$, the interbank liabilities $L_i$ and the equities $E_i$ of each $i$ of these banks, and we consider data for years 2008 and 2013 (i.e, during and after the global financial crisis) \citep{angelini2011interbank}.
We recall that the equity of a bank is the difference between its total positive positions and its total obligations to creditors. 
When the equity is positive the bank is solvent, otherwise it goes bankrupt (defaults) because it would not be able to refund its debts.
Since the chosen group of banks is not an isolated system, interbank assets and liabilities do not sum up to the same value. 
In order to have a closed system, we re-scale them such that $\sum_{j} A_j = \sum_{j} L_j$.

\subsection{Network Generation}

In the literature on financial networks, interbank markets are typically reconstructed from balance sheet data (before being tested for systemic risk) \citep{anand2017missing}. 
Here we use and generalize the approach of \cite{cimini2015systemic} to generate reconstructed financial networks (that is, compatible with balance sheet information) with different underlying topology. 
The method is grounded on statistical physics concepts applied to networks (see further details in \cite{squartini2018reconstruction,cimini2019statistical}). 

To create a single network instance, we first generate an unweighted directed graph by drawing each link $i \rightarrow j$ independently with probability:
\begin{equation}
p_{i \rightarrow j} = \frac{z A_i L_j}{1 + z A_i L_j},
\label{eq:pij}
\end{equation}
where $z \in (0, \infty)$ is a parameter that controls the density of the network.
Indeed, since the values of assets and liabilities are given, this probability is an increasing function of $z$, hence the link density of the network is proportional to the parameter $z$. 

After the link generation process, we assign a weight to each realized link as follows:
\begin{equation}
w_{i\rightarrow j}=\frac{A_i L_j}{\Omega p_{i\rightarrow j}}\,a_{i \rightarrow j}
\label{eq:wij}
\end{equation}
where the adjacency matrix element $a_{i \rightarrow j}$ equals 1 if the draw of Eq. (\ref{eq:pij}) was successful (and zero otherwise), and $\Omega=[(\sum_jA_j)(\sum_jL_j)]^{1/2}$. 
The final result is a weighted directed network given by the corresponding adjacency matrix $\mathrm{W}$, whose entries are the weights $\{w_{i \rightarrow j}\}$.
In the economic network literature this matrix is referred to as the asset matrix, while its transpose is called the liability matrix. 

Overall, this recipe is used to generate an ensemble of networks, having the property that interbank assets and liabilities of each bank are constrained in probability to their real values as 
$\langle\sum_jw_{i\rightarrow j}\rangle=A_i$ and $\langle\sum_jw_{j\rightarrow i}\rangle=L_i$ 
(here $\langle\cdot\rangle$ denotes the average over the ensemble).
Note that by using Eq. (\ref{eq:pij}) we allow the formation of self-loops in the network, because some of the top European banks do represent banking groups with an internal flow of money.
The alternative possibility would be to use the RAS algorithm to get rid of them, while preserving the imposed constraints \citep{squartini2017network}.

Importantly, the distribution of assets and liabilities across banks is heterogeneous, and with such an input our network construction method automatically generates a core-periphery structure, 
independently on the network density. In order to tune this outcome, we introduce a generalization of Eq. (\ref{eq:pij}):
\begin{equation}
p_{i\rightarrow j}=\frac{z\left(A_iL_j\right)^\phi}{1+z\left(A_i L_j\right)^\phi} \,, \quad \phi \in [0,1].
\label{eq:p'}
\end{equation}
The new parameter $\phi$ allows to model a wide range of network topologies (for fixed $z$), interpolating between the fitness-induced configuration model and the \ER random graph as the two limits $\phi = 1$ and $\phi = 0$, respectively.
Besides, weights assignment as for Eq. (\ref{eq:wij}) satisfies the constraints on interbank assets and liabilities of each bank whatever the choice of connection probabilities $\{p_{i \rightarrow j}\}$.

\subsection{Block Structure}

\begin{figure}
	\includegraphics[width = \textwidth]{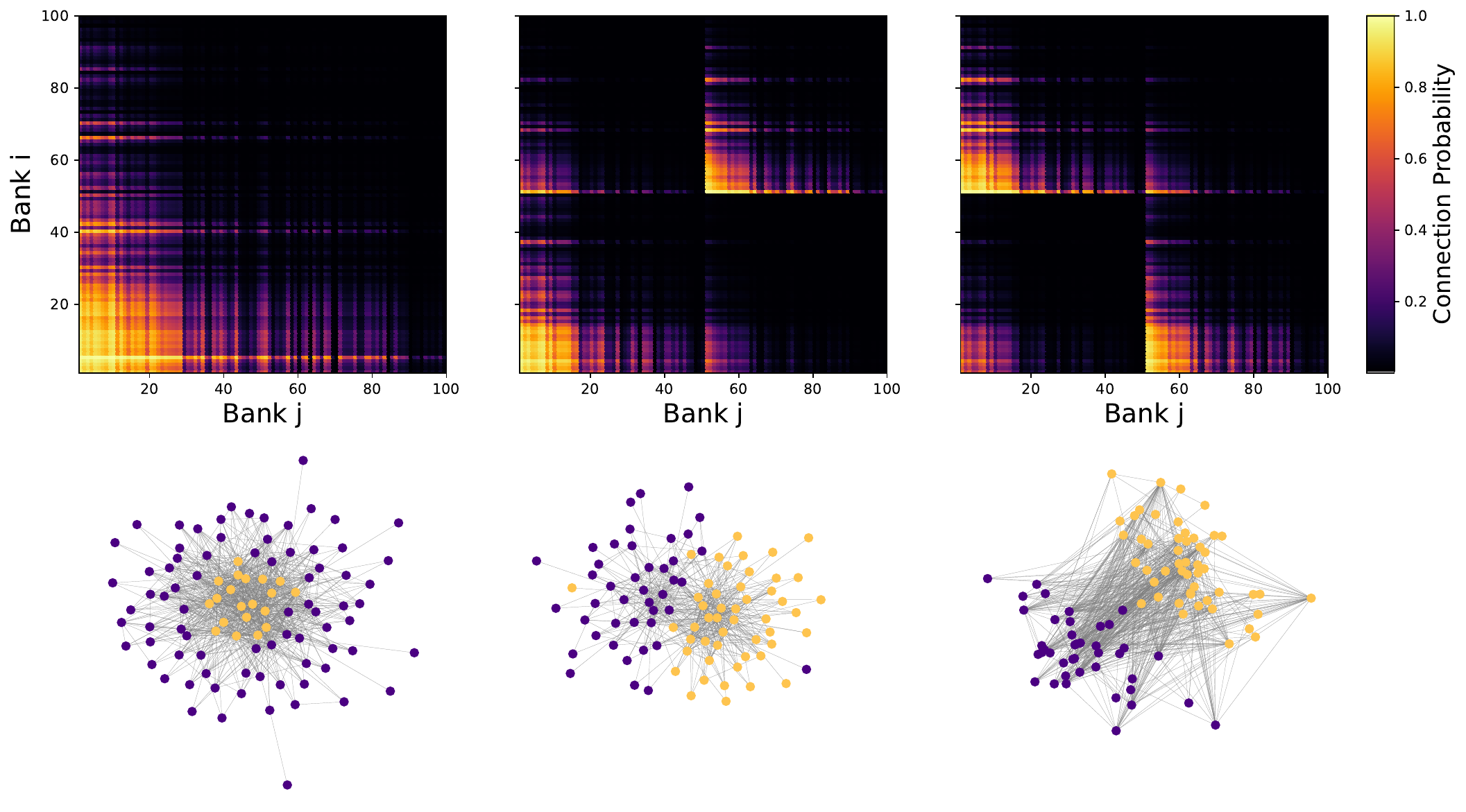}
    \caption{Examples of block structures obtained with the proposed models.
    \textit{Left:} core-periphery structure, generated using Eq. (\ref{eq:p'}) with $\phi = 1$.
    \textit{Middle:} modular assortative structure ($\phi = 1$, $\lambda = 0.1$).
    \textit{Right:} modular disassortative structure ($\phi = 1$, $\beta = 0.1$). For all cases we used $z = 10^{-9}$. 
    In the top row we report the heatmaps of the connection probability matrices (with yellow and purple representing denser and sparser regions of the network, respectively). In the bottom row we show the shape of a corresponding sample network. For the core-periphery case (left column), we display the core nodes (the top 20\% biggest banks) in yellow and the peripheral nodes in violet. For the other two cases the two colors refer to the two modules (of equal size).}
    \label{Fig:examples}
\end{figure}

The network reconstruction method just illustrated allows exploring different network structures. To this end, we can further decompose the adjacency matrix $\mathrm{W}$ into blocks. 
Here, for simplicity, we shall restrict our attention to the case of $\mathrm{W}$ having only four blocks:
$$\mathrm{W}= \begin{pmatrix} \mathrm{W}_{11} & \mathrm{W}_{12} \\ \mathrm{W}_{21} & \mathrm{W}_{22} \end{pmatrix} $$
Each block $\mathrm{W}_{nm}$, $n,m = \overline{1,2}$ represents a subgraph of the network in which the link density is characterized by $z_{nm}$, i.e, which is generated via Eqs (\ref{eq:pij})-(\ref{eq:wij}) using this $z_{nm}$. 
Among all possible topological block configurations, there are three distinct ones that we shall focus on, namely the core-periphery as well as the modular assortative and disassortative structures (see Figure \ref{Fig:examples}).

\paragraph*{Core-periphery topology} --- Of special interest in the investigation of financial networks is the core-periphery topology, in which there are two groups of banks (core and periphery) and a much higher link density within the first group than within the second group, and an intermediate link density between the two groups.
As mentioned in the last section, the network generated by Eq. (\ref{eq:pij}) inherently possesses a core-periphery structure.
Therefore, further using a parameterization of the form $z_{11} = z$, $z_{12} = z_{21} = \gamma z$, $z_{22} =\gamma^2 z$, where $\gamma \in [0,1]$, would already result in a core-periphery structure for $\gamma = 1$, with smaller values of $\gamma$ simply marginalising the peripheries.
This is why we only use Eq. (\ref{eq:p'}) to explore the transition between core-periphery and homogeneous topologies.

\paragraph*{Assortative modular topology} --- In this case the network is clustered into two groups of nodes (modules), with dense connections within the groups and sparse connections between them.
This configuration corresponds to the choice of $z_{11}$ and $z_{22}$ both much larger than $z_{12}$ and $z_{21}$.
Without loss of generality, we implement the assortative topology by setting $z_{11} = z_{22} = z$ and $z_{12} = z_{21}=\lambda z$, where $\lambda \in [0,1] $. 

\paragraph*{Disassortative modular topology} --- As an opposing configuration to the assortative modular structure, one can consider the case in which the interconnections between the two modules dominate over the connections inside each module.
The parameterization now is given by $z_{12} = z_{21} = z$ and $z_{11} = z_{22} = \beta z$, where $\beta \in [0,1]$.
Note that the limiting case $\beta\to0$ corresponds to a purely bipartite structure.\\

\noindent We remark that for both the assortative and disassortative topologies we consider in this paper, each block is generated using Eq. (\ref{eq:p'}) with the corresponding value of $z_{nm}$ and with $\phi = 1$.
As such, we have the signature of core-periphery structure within each module (as clearly visible in Figure~\ref{Fig:examples}.

\subsection{Shock propagation dynamics}

Once a network instance is constructed, we use the Furfine and DebtRank algorithms to model the propagation of shocks on top of it \citep{furfine2003interbank, bardoscia2015debtrank,bardoscia2016distress}.

\paragraph{Furfine algorithm}
The Furfine algorithm can be expressed entirely as a function of equities, interacting through the liabilities network and with a given value of external assets or liabilities.
The iterative map that represents the contagion dynamics is given by \citep{barucca2016network},

\begin{equation}
E_i(t+1)= e_i +\sum_{j=1}^{N}A_{ij}\left(\Theta(E_j(t)) + R\Theta(-E_j(t))\right) - L_i,
\label{eq:furfinedyn}
\end{equation}
where $R$ is the exogenous recovery rate, and $e_i$ is the external net balance, which can be determined by looking at the initial discrepancy between the equity values and the interbank net balance.

In particular, for the dataset under study, in case the initial equity is larger or smaller than the net balance of interbank assets and liabilities then it will imply the presence of a net external source of assets or liabilities, $e_i = E_i - A_i + L_i$.

From the iterative map, it is possible to define multiple measure of contagion losses, in particular we will focus on one measure of contagion loss that is the average relative equity loss, i.e. we assume a relative shock to our set of banks, and we consider the shocked values as our initial value of the equity vector. 
Starting from this new equity vector we identify the fixed point vector of the equity dynamics \eqref{eq:furfinedyn}, $E^*$.
The average relative equity loss is then given by:

\begin{equation}
\overline{E}_{loss}= \frac{\sum_i E^*_i - E_i(0)}{\sum_j E^{pre}_j}
\label{eq:Elossfurfine}
\end{equation}
where the losses are computed relatively to the initial pre-shock values, $E^{pre}_i$.

\paragraph{DebtRank}
Alternatively, for a given step $t$ of the dynamics, we consider the relative loss of equity $h_i(t)$ and the interbank leverage matrix $\Lambda_{ij}(t)$ of each bank $i$:
\begin{equation}
h_i(t)=\frac{E_i(0)-E_i(t)}{E_i(0)}
\label{eq:h_i(t)}
\end{equation}
\begin{equation}
\Lambda_{ij}(t)= \begin{cases} \dfrac{w_{i\to j}(0)}{E_i(0)} & \mbox{if bank $j$ has not defaulted up to time $(t-1)$} \\ ~~~0 & \mbox{otherwise} \mbox{~} \end{cases}
\label{lambdaij}
\end{equation}
where $E_i(0)$ is the initial equity of $i$. Assuming a loss given default of 100\%, the dynamical equation for the relative equity loss $h_{i}(t)$ reads:
\begin{equation}
h_i(t+1)=\min \left[ 1,h_i(t)+\sum_{j=1}^{N}\Lambda_{ij}(t)[p_j^D(t+1)-p_j^D(t)]\right]
\label{eq:h_i(t+1)}
\end{equation}
where $p_j^D(t)=h_j(t)e^{\alpha[h_j(t)-1]}$ is the probability of default of bank $j$ at step $t$. The controlling parameter $\alpha \in (0, \infty)$ allows switching continuously from the linear DebtRank ($\alpha=0$, meaning that the default probability is directly proportional to equity losses) \citep{bardoscia2015debtrank} to the Furfine algorithm ($\alpha\to\infty$, i.e., default occurs only when equity is depleted, and the bank is not contagious otherwise) \citep{furfine2003interbank}. 
The average relative equity loss at the end of the shock propagation dynamics $t^*$ is:
\begin{equation}
\overline{E}_{loss}=\sum_i \frac{[h_i(t^*)-h_i(1)]E_i(0)}{\sum_j E_j(0)}
\label{eq:Eloss}
\end{equation}
where $h_i(1)$ is the initial shock on $i$.
Hence $\overline{E}_{loss}$ does not account for the initial shock on the system, but only for the network effect on systemic losses.

We use two kinds of stopping conditions for simulations: either when the difference $\|[h(t)-h(t-1)]E(0)\|_2$ becomes smaller than a tolerance $tol = 10^{-5}$, or when the number of interactions is equal to $max_{iter}=10^5$.

\section{Results}

As mentioned before, our operative framework consists of building an ensemble of interbank networks (using balance sheet data for either years 2008 and 2013) and then using the DebtRank shock propagation dynamics to run stress tests on each of these networks. $\overline{E}_{loss}$ is the average outcome of the process over an ensemble of cardinality 100.

\begin{figure}
	\includegraphics[width = \textwidth]{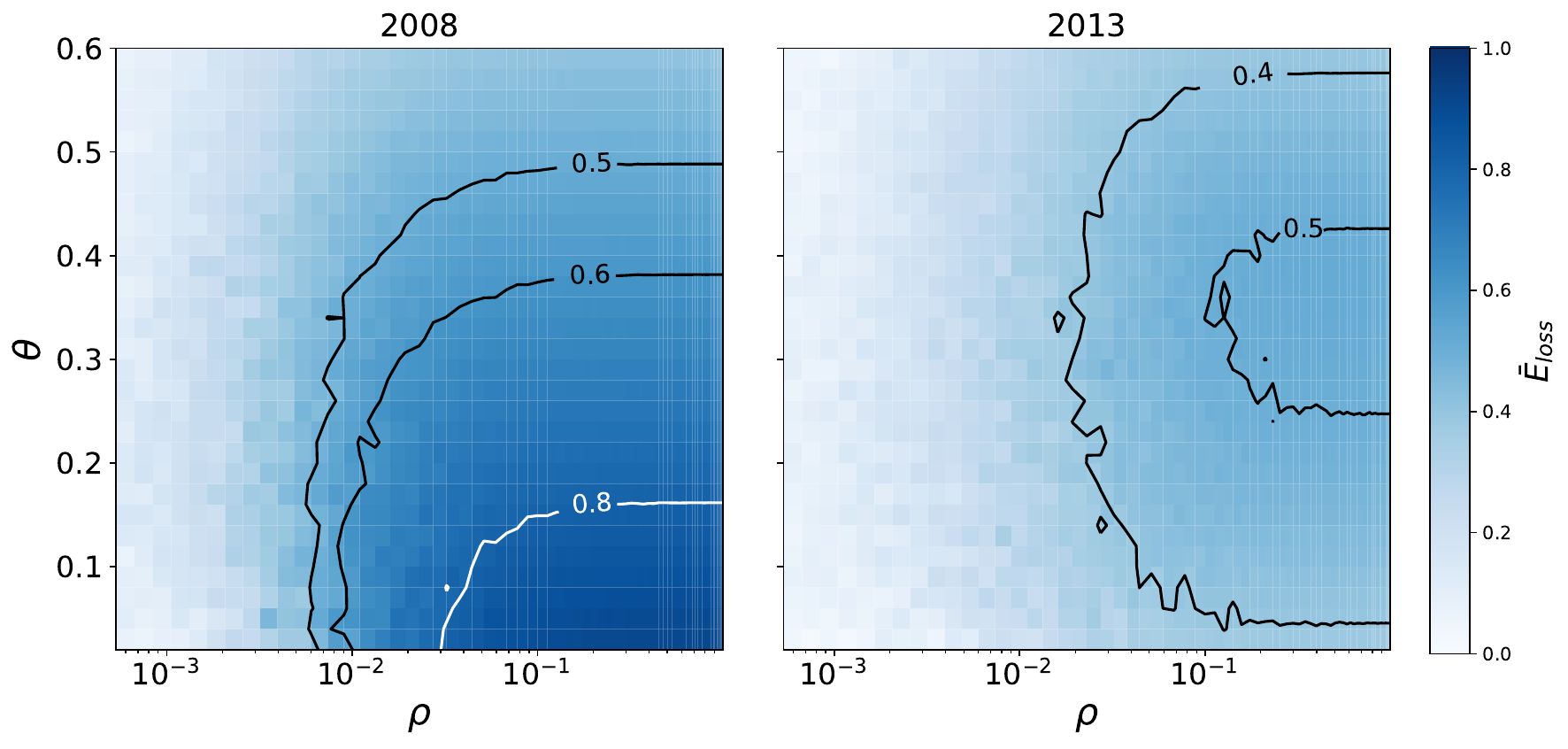}
    \caption{$\overline{E}_{loss}$ (computed with linear DebtRank) as a function of the link density and the magnitude of the uniform shock, for core-periphery networks built with data from 2008 (left) and 2013 (right) using Eq. (\ref{eq:p'}) with $\phi = 1$. 
    Darker (brighter) color refers to the higher (smaller) DebtRank, which corresponds to a more fragile (resilient) financial network.}
    \label{Fig:08-13}
\end{figure}

The first and most basic exercise we perform focuses on studying the standard fitness configuration model of \cite{cimini2015systemic}, generated by using Eq. (\ref{eq:p'}) with $\phi = 1$, and varying network density $\rho$.
In terms of initial shock, we consider a uniform shock by reducing the equity of each bank by a fraction $\theta$ of its initial value, which means $h_i(1)=\theta \, \, \forall i$.
Figure \ref{Fig:08-13} shows the result of this exercise for $\rho$ ranging from 0 to 1, and $\theta$ ranging from 0 to 0.6.
We see that $\overline{E}_{loss}$ increases monotonically with $\rho$.
This implies that the network becomes more fragile when it becomes more dense, consistently with the findings of \cite{bardoscia2017pathways}. 
In 2008 we find a very high value of $\overline{E}_{loss}$ also for very small $\theta$: network amplification effects are so important that they can wipe out the whole system also when the initial perturbation is minimal.
The decreasing of $\overline{E}_{loss}$ with $\theta$ is instead mainly due to the fact that initial shocks are not included in the computation of $\overline{E}_{loss}$, and indeed this quantity is bounded from above by $1-\theta$.
Finally, by comparing the 2008 data and the 2013 data, we find that the $\overline{E}_{loss}$ for every combination of $\rho$ and $\theta$ has substantially changed: the network is much more robust in 2013 than in 2008, especially for what concerns small shocks (even in the high density regime) \citep{cimini2016entangling}. 

\begin{figure}
	\includegraphics[width = \textwidth]{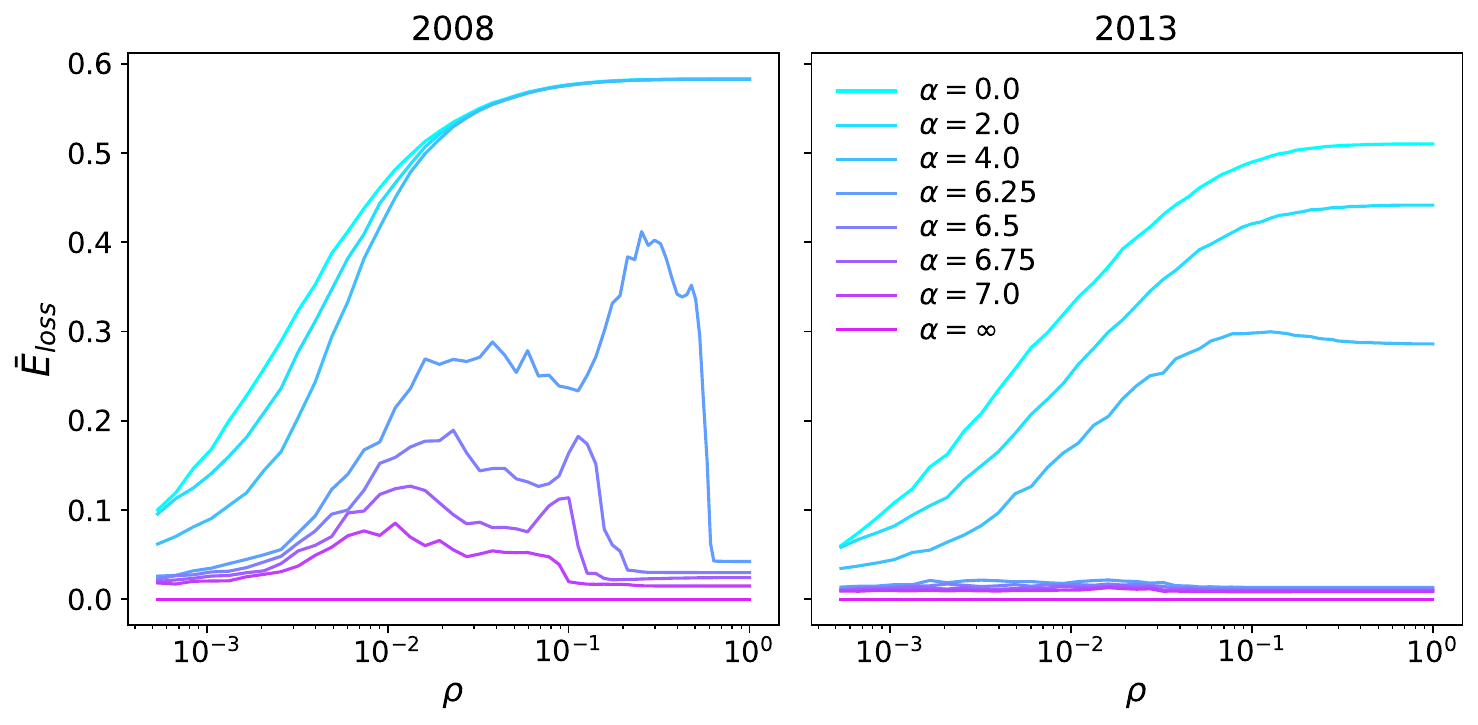}
      \caption{$\overline{E}_{loss}$ as a function of the link density for a fixed value $\theta = 0.4$ of uniform initial shock, for networks built with data from 2008 (left) and 2013 (right) using Eq. (\ref{eq:p'}) with $\phi = 1$. 
      Different curves correspond to different magnitudes of non-linearity in the DebtRank default probabilities entering in Eq. (\ref{eq:h_i(t+1)}).}
    \label{Fig:NLDR}
\end{figure}

Note that in this exercise we have looked at the case of linear DebtRank, which corresponds to the choice $\alpha = 0$ in Eq. (\ref{eq:h_i(t+1)}).
We now perform an analogous exercise but looking at different values of $\alpha$.
In particular, we are interested in the cases of DebtRank, Furfine, and the non-linear default probability in between.
To this end, we consider in Figure \ref{Fig:NLDR} the values $\alpha = 0, 2, 4, 6.25, 6.5, 6.75, 7, \infty$.
This set of parameters was chosen such that the reader can see a clear transition in $\overline{E}_{loss}$ behavior but at the same time the number of curves is small enough for the plot to be readable.
Concerning 2008 networks, in general for small $\alpha$ we see that $\overline{E}_{loss}$ increases with $\rho$, and converges towards the highest value of $\overline{E}_{loss}=1-\theta$. 
In contrast, for large $\alpha$ equity losses, due to the network, remain very small for both very dense and very sparse networks, and attain a maximum for intermediate density values. 
The transition between these two regimes appears around the case $\alpha = 5$, corresponding to a highly nonlinear default probability.
Moving to 2013 networks, the increasing behavior of $\overline{E}_{loss}$ with $\rho$ is generally observed also in this case, however $\overline{E}_{loss}$ for completely connected networks does depend on the value of $\alpha$.
The regime of nonlinear default probability shows very moderate losses, in line with the interpretation that in 2013 the interbank market was much more stable than in 2008.

\begin{figure}
	\includegraphics[width = \textwidth]{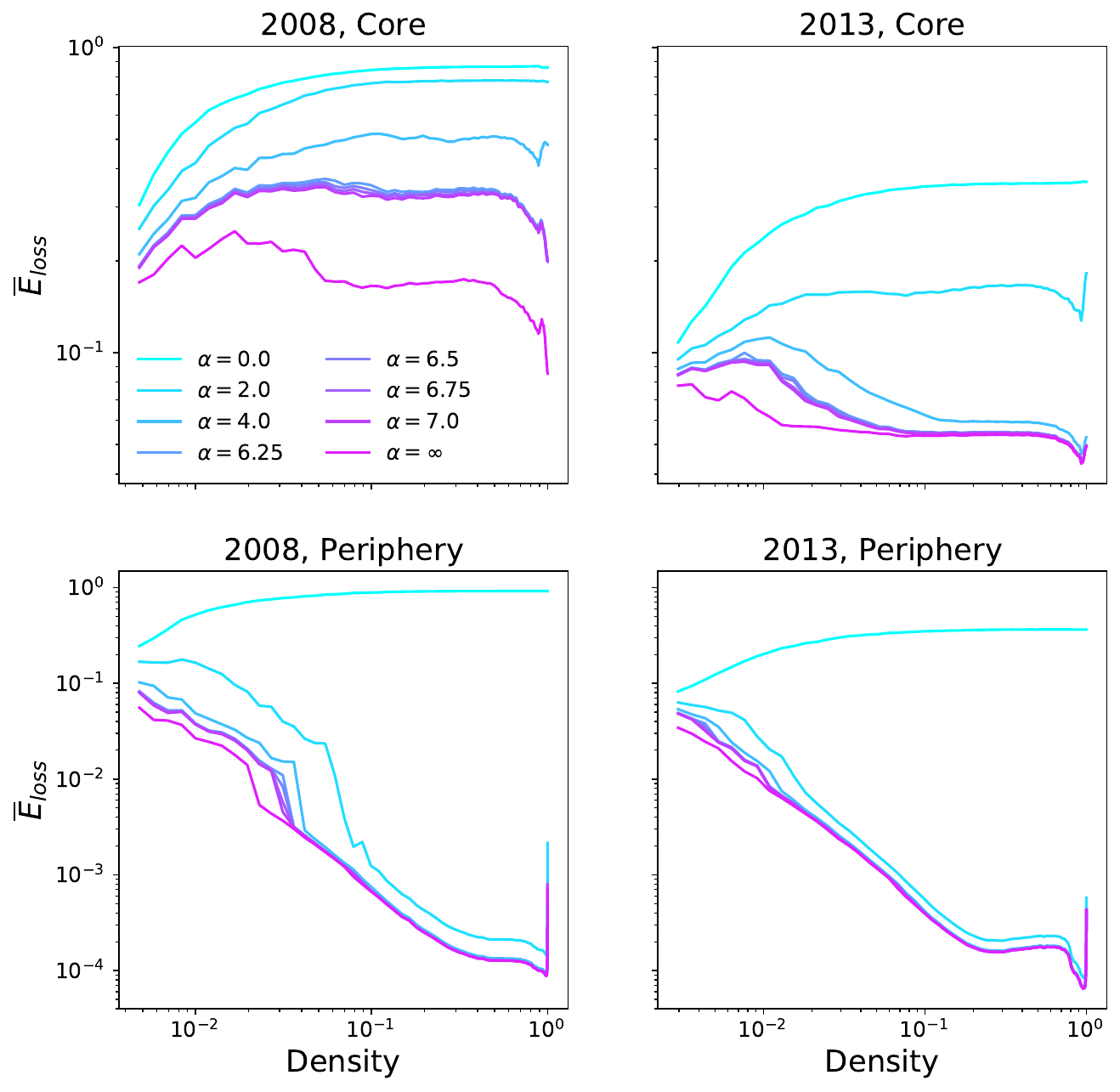}
      \caption{$\overline{E}_{loss}$ as a function of the link density for point initial shocks in the core (upper) or in the periphery (lower), for networks built with data from 2008 (left) and 2013 (right) using Eq. (\ref{eq:p'}) with $\phi = 1$. 
      Different curves correspond to different magnitudes of non-linearity in the default probabilities entering in Eq. (\ref{eq:h_i(t+1)}).}
    \label{Fig:PNLDR}
\end{figure}

\begin{figure}
	\includegraphics[width = \textwidth]{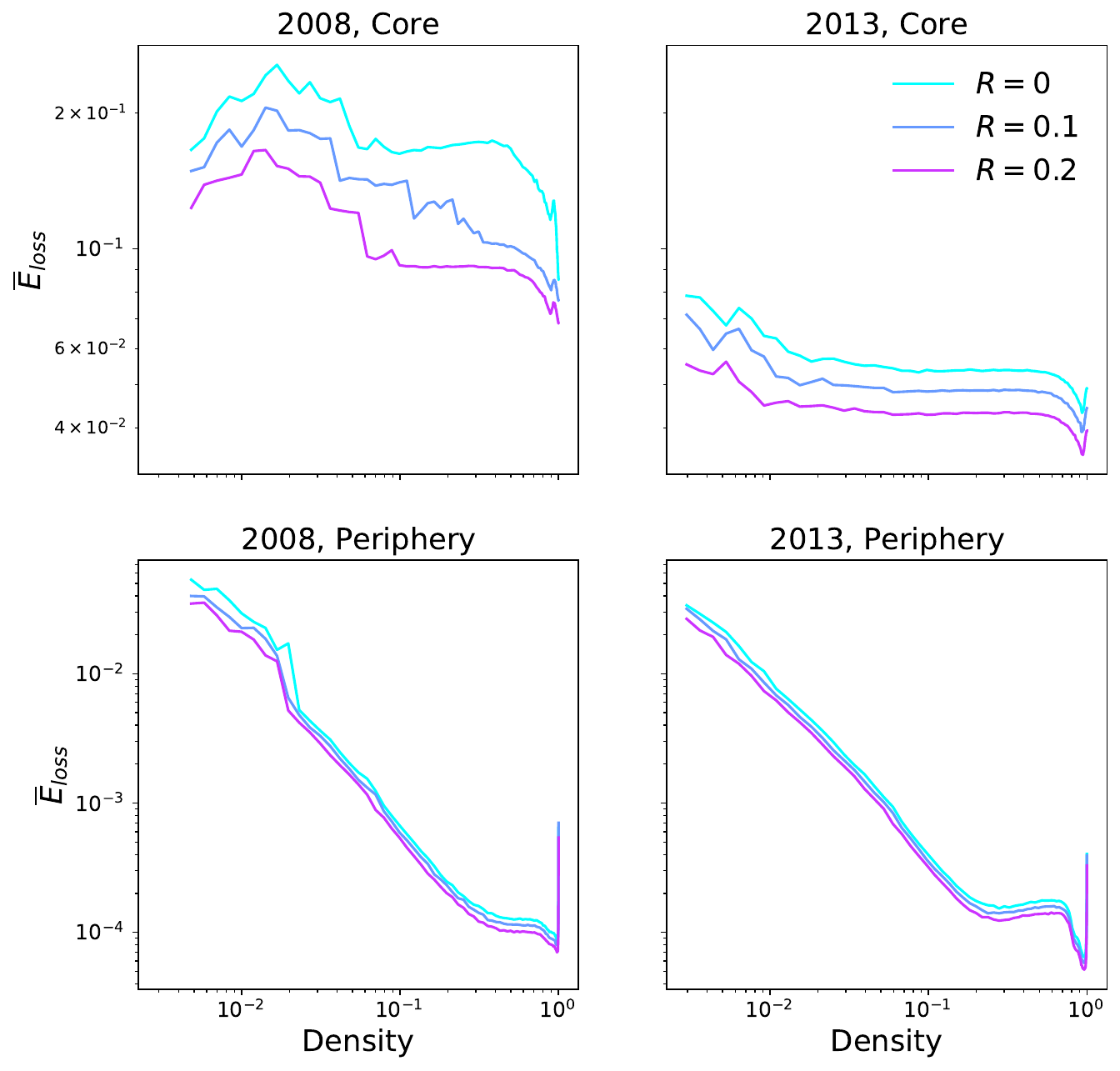}
      \caption{$\overline{E}_{loss}$ for the Furfine model as a function of the link density for point initial shocks in the core (upper) or in the periphery (lower), for networks built with data from 2008 (left) and 2013 (right) using Eq. (\ref{eq:p'}) with $\phi = 1$. 
      Different curves correspond to different values of the exogenous recovery rate R.}
    \label{Fig:recovery}
\end{figure}

The flat, and equal to zero, curve for the $\alpha = \infty$ case is a technical consequence of using homogeneous shocks with Furfine.
To address this issue, in addition to the uniform shock, we study the effect of defaulting a single bank either from core or from periphery.
In this latter case, we divide all the banks into core, middle and periphery (around 30 banks in each group) by ordering them according to their reconstructed degree, i.e., to their exposure (as a consequence of the fitness ansatz used in the reconstruction procedure).
Then we choose to default one of them randomly and average the result over 100 picks from one given group.
As shown in Figure \ref{Fig:PNLDR}, the behavior of $\overline{E}_{loss}$ is significantly different for core and periphery shocks.
The latter results in either increase of the risk as a function of density for $\alpha = 0$ (linear case), or decrease when $\alpha > 0$.
When shocking the core, on the other hand, we observe an increase of risk for small densities, but at some point it reaches a maximum and then drops (non-linear case) or remains at the same level (linear case).
Additionally, Figure \ref{Fig:recovery} shows how this dependence is affected by the recovery rate $R$ in the Furfine case.
Obviously the losses are decreased with increasing $R$, however, the overall function form does not change.
Although qualitatively 2008 and 2013 are very similar, 2013 is more robust for all analysed cases.

\begin{figure}
	\includegraphics[width = \textwidth]{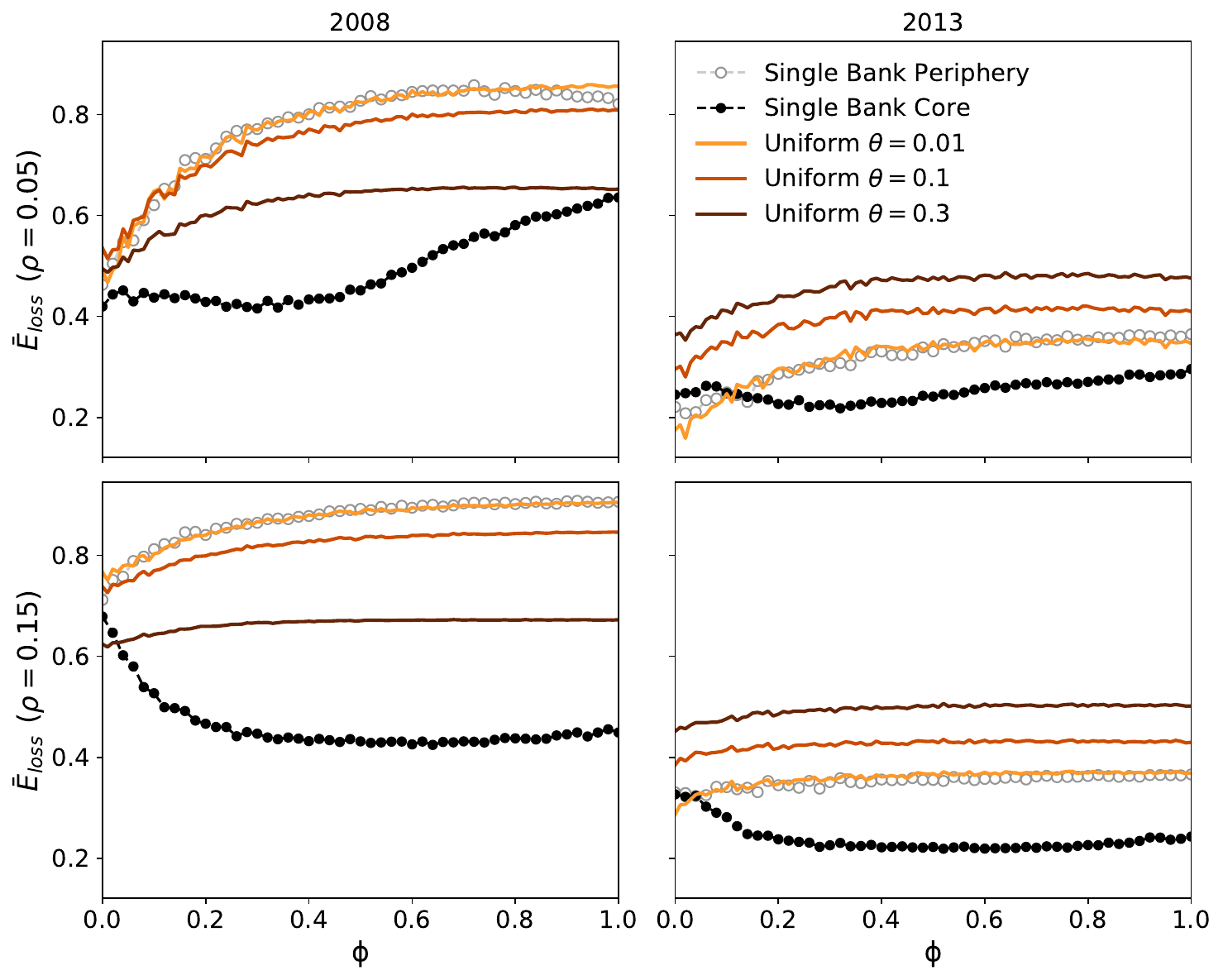}
    \caption{$\overline{E}_{loss}$ (computed with linear DebtRank) as a function of the parameter $\phi$ of Eq. (\ref{eq:p'}) tuning the strength of the core-periphery structure, for both 2008 and 2013 interbank networks. 
    We consider fixed values of the density $\rho = 0.05$ (left) and $\rho = 0.15$ (right), and different initial shocks: either 
    uniform for all banks, or consisting of a single initial default (in the core or in the periphery).}
    \label{Fig:CP}
\end{figure}

Up to this point we have considered core-periphery interbank network structures generated using Eq. (\ref{eq:p'}) with $\phi = 1$.
We now consider other values of $\phi$, up to the case $\phi = 0$ correpsonding to an \ER topology.
Figure \ref{Fig:CP} shows network losses $\overline{E}_{loss}$ as a function of $\phi$ and for different types of initial shocks.
We see that uniform shocks cause a loss which is increasing with $\phi$, so the more core-periphery the structure, the more fragile the system is.
This behavior is observed for different values of the network density and for both 2008 and 2013.
Moreover, as we already observed in Figure \ref{Fig:08-13}, in 2008 small uniform shocks cause higher relative losses than bigger ones, whereas in 2013 the opposite is true.
One should note, however, that if we sum the relative loss and the initial loss, such sum is always higher for higher initial shock.
Additionally, 2013 is much more robust with $\overline{E}_{loss}$ not exceeding $0.5$ even for $\phi = 1.0$ and $\theta = 0.3$.

Interestingly, the behavior changes when we consider single bank defaults.
When the network is more random ($\phi\sim0$) both types of shocks result in a similar value of $\overline{E}_{loss}$. 
Yet as the distinction between core and periphery emerges and becomes more marked, the result of shocking either of them changes.
As can be seen in all four panels of figure \ref{Fig:CP}, shocking the periphery is almost equivalent to the small uniform shock scenario.
Indeed, the average initial shock corresponding to single bank default in the periphery is around 0.1\% of the whole initial equity.
On the other hand, the loss induced by defaulting a bank belonging to the core is not an increasing function of $\phi$.
For denser networks, with $\rho = 0.15$, it decreases in the near random regime and then it stays constant.
In the case of $\rho = 0.05$, $\overline{E}_{loss}$ reaches a minimum and then it increases with $\phi$.
The existence of an optimal $\phi$, from the core defaults perspective, is an important observation from the regulatory point of view.
Note that the average initial loss corresponding to the default of a core bank is more than 4\% of the entire system equity.

\begin{figure}
	\includegraphics[width = \textwidth]{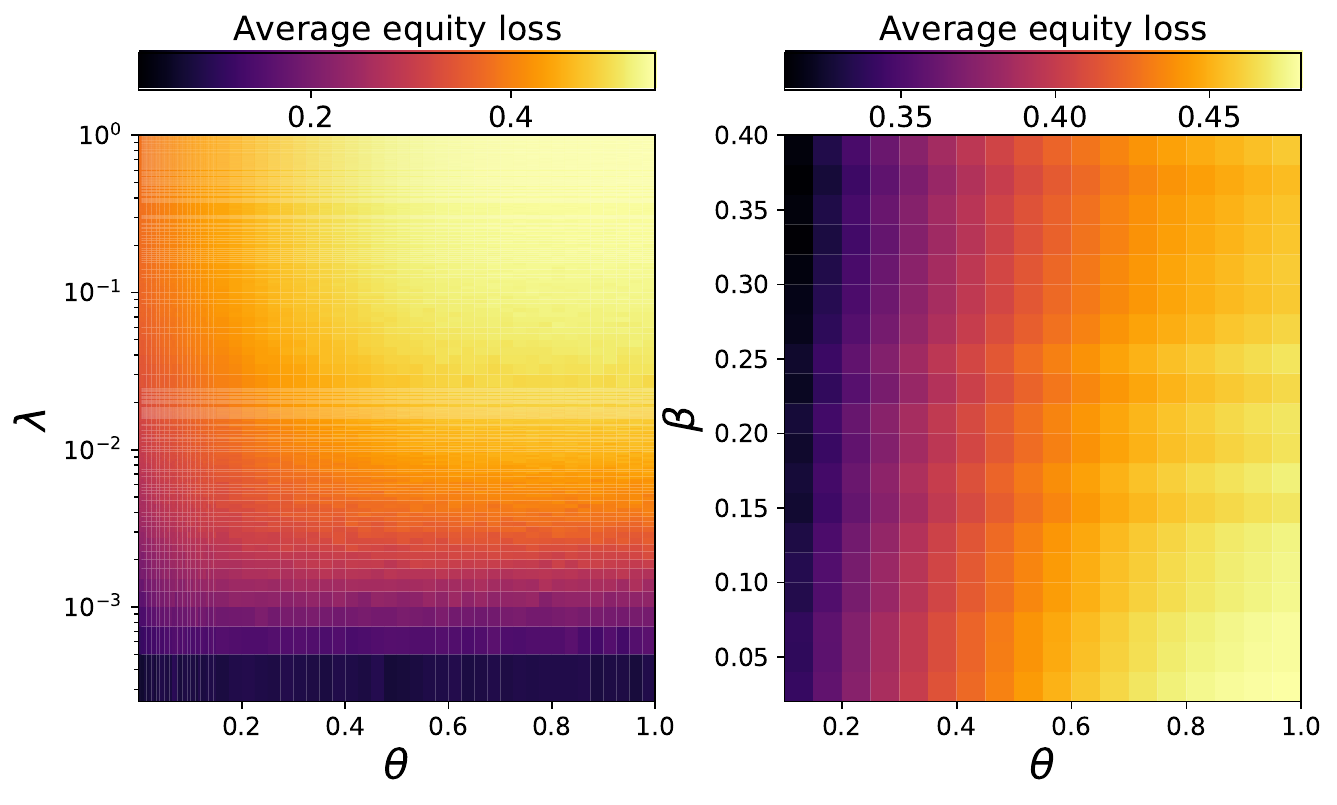}
    \caption{$\overline{E}_{loss}$ (computed with linear DebtRank) as a function of the uniform initial shock $\theta$ and model parameter $\lambda$ and $\beta$ of the modular assortative (left) and disassortative (right) topological structure, respectively. In both cases $\phi = 1$ and $\rho = 0.1$, and the interbank networks refer to year 2008.}
    \label{Fig:ComBip}
\end{figure}

We finally consider the modular assortative and disassortative topologies, by varying the structural parameters $\lambda$ and $\beta$ but for fixed $\phi=1$.
Figure \ref{Fig:ComBip} shows the results of an exercise in which we uniformly shock banks from the first module and measure $\overline{E}_{loss}$ for the second module.
We observe that the assortative structure can be quite resilient if the different blocks are scarcely connected.
On the other hand, above a given value of $\lambda$ the structure becomes almost as fragile as in the case without the blocks ($\lambda = 1$).
Concerning disassortative structures, systemic risk decreases when we move away from a pure bipartite structure.
The differences, however, are relatively small and there is no jump similar to the one observed for assortative structures.
For a constant density, this may be a result of decreasing number of connections between the two groups when we increase $\beta$.
In this way the losses are not transmitted to the other side as quickly as in the purely bipartite case.
Qualitatively similar results were obtained for the 2013 data.

\section{Discussion}

In this work we have examined different structures of the interbank network and have shown how they affect the systemic risk.
In addition, we also used a variety of shock types and changed the way they propagate across the network.
These results provide additional evidence of how complex the interbank system is and how many variables are involved in determining its resilience.

In the simplest situation of a single-block network with a core-periphery structure, systemic risk monotonically increases with the density, but there is a quantitative difference between the behavior observed in 2008 and 2013: the crisis (and the consequent regulatory interventions) shaped the banks balances in a way that afterwards the interbank market became much more robust to small shocks even in the high density regime.

Analysis of different shapes of the bank default probability further shows the differences between pre and post crisis.
In 2008 there are basically two regimes of the shocks propagation, depending on the network density and the parameter $\alpha$---which can be also seen as the amount of confidence market participants have in the ability of counterparts to recover from equity losses.
Indeed if the confidence in the system is not high enough, systemic losses become widespread, otherwise they remain small.
On the other hand in 2013, even for low confidence levels, increasing the density does not lead to overall losses equal to what is observed for the linear case (that corresponding to the lowest level of trust in the counterpart).
Note that these results were obtained with uniform shocks over all banks.
As shown in \cite{mastromatteo2012reconstruction}, if we consider the default of single banks as initial shocks, we expect that increasing the density could help the system to withstand the shock.

As a matter of fact, the core-periphery structure is naturally generated by the network (re)construction method described by Eq. (\ref{eq:pij}) when assets and liabilities are having fat-tailed distribution.
In this scenario, the only topological parameter that can be tuned is the network density, which allows to switch from networks with a few contracts of large amounts to networks with many contracts of small amounts.
However, by introducing the parameter $\phi$ in Eq. (\ref{eq:p'}) we were also able to continuously change the network structure from a random one ($\phi = 0$) to a core-periphery one ($\phi = 1$).
We then found that the core periphery structure is less resilient, at least for uniform shocks.
This confirms the well-known observation that strongly connected nodes enhance the shock propagation \citep{huser2015too}.

For the dataset under study default contagion algorithms, i.e. Furfine or DebtRank for $\alpha \rightarrow \infty$, would not yield any equity losses for uniform shocks, i.e. no propagation occurs unless a bank actually defaults.
On the other hand, distress contagion algorithms, DebtRank for finite $\alpha$, distinguishes between the vulnerabilities of different institutions also before any default occurs.

In the case of an initial bank defaults, the propagation depends on whether the defaults appeared in the core or in the peripheries.
The peripheries are clearly more fragile as the structure becomes more core-periphery than random.
Importantly, the core is quite robust, especially in between the two extreme structures.
In this case, both default and distress contagion algorithms display equity losses. 

In the last step of our analysis we looked at the block structure of the network.
In particular, an assortative modular structure can well represent a market of several countries, in which home and foreign connectivities are different (the former being typically much larger).
On the other hand, the disassortative modular case is often observed in financial networks, especially at low data aggregation scale for which a bank is either a lender or a borrower but not both \citep{barucca2016disentangling}.
In both cases we showed how a shock originating in one block propagates to the other.
For the assortative case we observe a significant jump of systemic risk above some given density of connections between the blocks.
On the other hand, the disassortative structure does not reveal any kind of similar discontinuity.
Nevertheless, moving away from the pure bipartite case slowly decreases the systemic risk.

Overall, we showed that the outcome of a systemic event is very much dependent on the details of both the underlying network and the shock propagation. 
We believe this observation is relevant also for more general frameworks such as multilayer financial networks \citep{battiston2018financial,podelna2015multilayer,bargigli2015multiplex} and networks of portfolio holdings \citep{caccioli2014stability,cont2016fire,gualdi2016statistically,greenwood2015vulnerable,pichler2018systemic}.
Our results can thus be of inspiration for more in-depth analyses and may provide useful insights to regulators in trying to shape a more resilient financial system. 
In the latter case, the knowledge about the systemic consequences of changing the interbank network density or re-shaping its structure, would be a valuable argument for or against a given solution.
Moreover, studying the effects caused by different shock propagation types, can help predicting the outcomes of different stress scenarios.

\section*{Acknowledgements}

This work is the output of the first Complexity72h workshop, held at IMT School in Lucca, 7-11 May 2018. https://complexity72h.weebly.com/

\bibliographystyle{apacite}
\bibliography{bibliography}

\begin{thebibliography}{}

\bibitem [\protect \citeauthoryear {%
Acemoglu%
, Ozdaglar%
\BCBL {}\ \BBA {} Tahbaz-Salehi%
}{%
Acemoglu%
\ \protect \BOthers {.}}{%
{\protect \APACyear {2015}}%
}]{%
acemoglu2015systemic}
\APACinsertmetastar {%
acemoglu2015systemic}%
\begin{APACrefauthors}%
Acemoglu, D.%
, Ozdaglar, A.%
\BCBL {}\ \BBA {} Tahbaz-Salehi, A.%
\end{APACrefauthors}%
\unskip\
\newblock
\APACrefYearMonthDay{2015}{}{}.
\newblock
{\BBOQ}\APACrefatitle {Systemic risk and stability in financial networks}
  {Systemic risk and stability in financial networks}.{\BBCQ}
\newblock
\APACjournalVolNumPages{American Economic Review}{105}{2}{564--608}.
\newblock
\begin{APACrefDOI} \doi{10.1257/aer.20130456} \end{APACrefDOI}
\PrintBackRefs{\CurrentBib}

\bibitem [\protect \citeauthoryear {%
Acharya%
, Pedersen%
, Philippon%
\BCBL {}\ \BBA {} Richardson%
}{%
Acharya%
\ \protect \BOthers {.}}{%
{\protect \APACyear {2017}}%
}]{%
acharya2017measuring}
\APACinsertmetastar {%
acharya2017measuring}%
\begin{APACrefauthors}%
Acharya, V\BPBI V.%
, Pedersen, L\BPBI H.%
, Philippon, T.%
\BCBL {}\ \BBA {} Richardson, M.%
\end{APACrefauthors}%
\unskip\
\newblock
\APACrefYearMonthDay{2017}{}{}.
\newblock
{\BBOQ}\APACrefatitle {Measuring systemic risk} {Measuring systemic
  risk}.{\BBCQ}
\newblock
\APACjournalVolNumPages{The Review of Financial Studies}{30}{1}{2--47}.
\newblock
\begin{APACrefDOI} \doi{10.1093/rfs/hhw088} \end{APACrefDOI}
\PrintBackRefs{\CurrentBib}

\bibitem [\protect \citeauthoryear {%
Allen%
\ \BBA {} Gale%
}{%
Allen%
\ \BBA {} Gale%
}{%
{\protect \APACyear {2007}}%
}]{%
allen2007systemic}
\APACinsertmetastar {%
allen2007systemic}%
\begin{APACrefauthors}%
Allen, F.%
\BCBT {}\ \BBA {} Gale, D.%
\end{APACrefauthors}%
\unskip\
\newblock
\APACrefYearMonthDay{2007}{}{}.
\newblock
{\BBOQ}\APACrefatitle {Systemic Risk and Regulation} {Systemic risk and
  regulation}.{\BBCQ}
\newblock
\BIn{} \APACrefbtitle {The risks of financial institutions} {The risks of
  financial institutions}\ (\BPGS\ 341--376).
\newblock
\APACaddressPublisher{}{University of Chicago Press}.
\PrintBackRefs{\CurrentBib}

\bibitem [\protect \citeauthoryear {%
Allen%
, Hryckiewicz%
, Kowalewski%
\BCBL {}\ \BBA {} T\"umer-Alkan%
}{%
Allen%
\ \protect \BOthers {.}}{%
{\protect \APACyear {2014}}%
}]{%
allen2014transmission}
\APACinsertmetastar {%
allen2014transmission}%
\begin{APACrefauthors}%
Allen, F.%
, Hryckiewicz, A.%
, Kowalewski, O.%
\BCBL {}\ \BBA {} T\"umer-Alkan, G.%
\end{APACrefauthors}%
\unskip\
\newblock
\APACrefYearMonthDay{2014}{}{}.
\newblock
{\BBOQ}\APACrefatitle {Transmission of financial shocks in loan and deposit
  markets: Role of interbank borrowing and market monitoring} {Transmission of
  financial shocks in loan and deposit markets: Role of interbank borrowing and
  market monitoring}.{\BBCQ}
\newblock
\APACjournalVolNumPages{Journal of Financial Stability}{15}{}{112--126}.
\newblock
\begin{APACrefDOI} \doi{10.1016/j.jfs.2014.09.005} \end{APACrefDOI}
\PrintBackRefs{\CurrentBib}

\bibitem [\protect \citeauthoryear {%
Amini%
, Cont%
\BCBL {}\ \BBA {} Minca%
}{%
Amini%
\ \protect \BOthers {.}}{%
{\protect \APACyear {2016}}%
}]{%
amini2016resilience}
\APACinsertmetastar {%
amini2016resilience}%
\begin{APACrefauthors}%
Amini, H.%
, Cont, R.%
\BCBL {}\ \BBA {} Minca, A.%
\end{APACrefauthors}%
\unskip\
\newblock
\APACrefYearMonthDay{2016}{}{}.
\newblock
{\BBOQ}\APACrefatitle {Resilience to contagion in financial networks}
  {Resilience to contagion in financial networks}.{\BBCQ}
\newblock
\APACjournalVolNumPages{Mathematical Finance}{26}{2}{329--365}.
\newblock
\begin{APACrefDOI} \doi{10.1111/mafi.12051} \end{APACrefDOI}
\PrintBackRefs{\CurrentBib}

\bibitem [\protect \citeauthoryear {%
Anand%
\ \protect \BOthers {.}}{%
Anand%
\ \protect \BOthers {.}}{%
{\protect \APACyear {2018}}%
}]{%
anand2017missing}
\APACinsertmetastar {%
anand2017missing}%
\begin{APACrefauthors}%
Anand, K.%
, van Lelyveld, I.%
, Banai, A.%
, Christiano~Silva, T.%
, Friedrich, S.%
, Garratt, R.%
\BDBL {}Silvestri, L.%
\end{APACrefauthors}%
\unskip\
\newblock
\APACrefYearMonthDay{2018}{}{}.
\newblock
{\BBOQ}\APACrefatitle {The missing links: A global study on uncovering
  financial network structure from partial data} {The missing links: A global
  study on uncovering financial network structure from partial data}.{\BBCQ}
\newblock
\APACjournalVolNumPages{Journal of Financial Stability}{35}{}{107--119}.
\newblock
\begin{APACrefDOI} \doi{10.1016/j.jfs.2017.05.012} \end{APACrefDOI}
\PrintBackRefs{\CurrentBib}

\bibitem [\protect \citeauthoryear {%
Angelini%
, Nobili%
\BCBL {}\ \BBA {} Picillo%
}{%
Angelini%
\ \protect \BOthers {.}}{%
{\protect \APACyear {2011}}%
}]{%
angelini2011interbank}
\APACinsertmetastar {%
angelini2011interbank}%
\begin{APACrefauthors}%
Angelini, P.%
, Nobili, A.%
\BCBL {}\ \BBA {} Picillo, C.%
\end{APACrefauthors}%
\unskip\
\newblock
\APACrefYearMonthDay{2011}{}{}.
\newblock
{\BBOQ}\APACrefatitle {The interbank market after August 2007: What has
  changed, and why?} {The interbank market after august 2007: What has changed,
  and why?}{\BBCQ}
\newblock
\APACjournalVolNumPages{Journal of Money, Credit and Banking}{43}{5}{923-958}.
\newblock
\begin{APACrefDOI} \doi{10.1111/j.1538-4616.2011.00402.x} \end{APACrefDOI}
\PrintBackRefs{\CurrentBib}

\bibitem [\protect \citeauthoryear {%
Bardoscia%
, Barucca%
, Codd%
\BCBL {}\ \BBA {} Hill%
}{%
Bardoscia%
\ \protect \BOthers {.}}{%
{\protect \APACyear {2019}}%
}]{%
bardoscia2019forward}
\APACinsertmetastar {%
bardoscia2019forward}%
\begin{APACrefauthors}%
Bardoscia, M.%
, Barucca, P.%
, Codd, A\BPBI B.%
\BCBL {}\ \BBA {} Hill, J.%
\end{APACrefauthors}%
\unskip\
\newblock
\APACrefYearMonthDay{2019}{}{}.
\newblock
{\BBOQ}\APACrefatitle {Forward-looking solvency contagion} {Forward-looking
  solvency contagion}.{\BBCQ}
\newblock
\APACjournalVolNumPages{Journal of Economic Dynamics and
  Control}{108}{}{103755}.
\PrintBackRefs{\CurrentBib}

\bibitem [\protect \citeauthoryear {%
Bardoscia%
, Battiston%
, Caccioli%
\BCBL {}\ \BBA {} Caldarelli%
}{%
Bardoscia%
\ \protect \BOthers {.}}{%
{\protect \APACyear {2015}}%
}]{%
bardoscia2015debtrank}
\APACinsertmetastar {%
bardoscia2015debtrank}%
\begin{APACrefauthors}%
Bardoscia, M.%
, Battiston, S.%
, Caccioli, F.%
\BCBL {}\ \BBA {} Caldarelli, G.%
\end{APACrefauthors}%
\unskip\
\newblock
\APACrefYearMonthDay{2015}{}{}.
\newblock
{\BBOQ}\APACrefatitle {DebtRank: A microscopic foundation for shock
  propagation} {Debtrank: A microscopic foundation for shock
  propagation}.{\BBCQ}
\newblock
\APACjournalVolNumPages{PLoS ONE}{10}{6}{e0130406}.
\newblock
\begin{APACrefDOI} \doi{10.1371/journal.pone.0130406} \end{APACrefDOI}
\PrintBackRefs{\CurrentBib}

\bibitem [\protect \citeauthoryear {%
Bardoscia%
, Battiston%
, Caccioli%
\BCBL {}\ \BBA {} Caldarelli%
}{%
Bardoscia%
\ \protect \BOthers {.}}{%
{\protect \APACyear {2017}}%
}]{%
bardoscia2017pathways}
\APACinsertmetastar {%
bardoscia2017pathways}%
\begin{APACrefauthors}%
Bardoscia, M.%
, Battiston, S.%
, Caccioli, F.%
\BCBL {}\ \BBA {} Caldarelli, G.%
\end{APACrefauthors}%
\unskip\
\newblock
\APACrefYearMonthDay{2017}{}{}.
\newblock
{\BBOQ}\APACrefatitle {Pathways towards instability in financial networks}
  {Pathways towards instability in financial networks}.{\BBCQ}
\newblock
\APACjournalVolNumPages{Nature Communications}{8}{}{14416}.
\newblock
\begin{APACrefDOI} \doi{10.1038/ncomms14416} \end{APACrefDOI}
\PrintBackRefs{\CurrentBib}

\bibitem [\protect \citeauthoryear {%
Bardoscia%
, Caccioli%
, Perotti%
, Vivaldo%
\BCBL {}\ \BBA {} Caldarelli%
}{%
Bardoscia%
\ \protect \BOthers {.}}{%
{\protect \APACyear {2016}}%
}]{%
bardoscia2016distress}
\APACinsertmetastar {%
bardoscia2016distress}%
\begin{APACrefauthors}%
Bardoscia, M.%
, Caccioli, F.%
, Perotti, J\BPBI I.%
, Vivaldo, G.%
\BCBL {}\ \BBA {} Caldarelli, G.%
\end{APACrefauthors}%
\unskip\
\newblock
\APACrefYearMonthDay{2016}{}{}.
\newblock
{\BBOQ}\APACrefatitle {Distress propagation in complex networks: The case of
  non-linear DebtRank} {Distress propagation in complex networks: The case of
  non-linear debtrank}.{\BBCQ}
\newblock
\APACjournalVolNumPages{PLoS ONE}{11}{10}{e0163825}.
\newblock
\begin{APACrefDOI} \doi{10.1371/journal.pone.0163825} \end{APACrefDOI}
\PrintBackRefs{\CurrentBib}

\bibitem [\protect \citeauthoryear {%
Bargigli%
, di Iasio%
, Infante%
, Lillo%
\BCBL {}\ \BBA {} Pierobon%
}{%
Bargigli%
\ \protect \BOthers {.}}{%
{\protect \APACyear {2015}}%
}]{%
bargigli2015multiplex}
\APACinsertmetastar {%
bargigli2015multiplex}%
\begin{APACrefauthors}%
Bargigli, L.%
, di Iasio, G.%
, Infante, L.%
, Lillo, F.%
\BCBL {}\ \BBA {} Pierobon, F.%
\end{APACrefauthors}%
\unskip\
\newblock
\APACrefYearMonthDay{2015}{}{}.
\newblock
{\BBOQ}\APACrefatitle {The multiplex structure of interbank networks} {The
  multiplex structure of interbank networks}.{\BBCQ}
\newblock
\APACjournalVolNumPages{Quantitative Finance}{15}{4}{673--691}.
\newblock
\begin{APACrefDOI} \doi{10.1080/14697688.2014.968356} \end{APACrefDOI}
\PrintBackRefs{\CurrentBib}

\bibitem [\protect \citeauthoryear {%
Barucca%
\ \protect \BOthers {.}}{%
Barucca%
\ \protect \BOthers {.}}{%
{\protect \APACyear {2016}}%
}]{%
barucca2016network}
\APACinsertmetastar {%
barucca2016network}%
\begin{APACrefauthors}%
Barucca, P.%
, Bardoscia, M.%
, Caccioli, F.%
, D'Errico, M.%
, Visentin, G.%
, Battiston, S.%
\BCBL {}\ \BBA {} Caldarelli, G.%
\end{APACrefauthors}%
\unskip\
\newblock
\APACrefYearMonthDay{2016}{}{}.
\newblock
\APACrefbtitle {Network valuation in financial systems.} {Network valuation in
  financial systems.}
\newblock
\APAChowpublished {\url{https://arxiv.org/abs/1606.05164}}.
\PrintBackRefs{\CurrentBib}

\bibitem [\protect \citeauthoryear {%
Barucca%
\ \BBA {} Lillo%
}{%
Barucca%
\ \BBA {} Lillo%
}{%
{\protect \APACyear {2016}}%
}]{%
barucca2016disentangling}
\APACinsertmetastar {%
barucca2016disentangling}%
\begin{APACrefauthors}%
Barucca, P.%
\BCBT {}\ \BBA {} Lillo, F.%
\end{APACrefauthors}%
\unskip\
\newblock
\APACrefYearMonthDay{2016}{}{}.
\newblock
{\BBOQ}\APACrefatitle {Disentangling bipartite and core-periphery structure in
  financial networks} {Disentangling bipartite and core-periphery structure in
  financial networks}.{\BBCQ}
\newblock
\APACjournalVolNumPages{Chaos, Solitons \& Fractals}{88}{}{244-253}.
\newblock
\begin{APACrefDOI} \doi{10.1016/j.chaos.2016.02.004} \end{APACrefDOI}
\PrintBackRefs{\CurrentBib}

\bibitem [\protect \citeauthoryear {%
Battiston%
, Caldarelli%
, D’Errico%
\BCBL {}\ \BBA {} Gurciullo%
}{%
Battiston%
, Caldarelli%
\BCBL {}\ \protect \BOthers {.}}{%
{\protect \APACyear {2016}}%
}]{%
battiston2016leveraging}
\APACinsertmetastar {%
battiston2016leveraging}%
\begin{APACrefauthors}%
Battiston, S.%
, Caldarelli, G.%
, D’Errico, M.%
\BCBL {}\ \BBA {} Gurciullo, S.%
\end{APACrefauthors}%
\unskip\
\newblock
\APACrefYearMonthDay{2016}{}{}.
\newblock
{\BBOQ}\APACrefatitle {Leveraging the network: A stress-test framework based on
  DebtRank} {Leveraging the network: A stress-test framework based on
  debtrank}.{\BBCQ}
\newblock
\APACjournalVolNumPages{Statistics \& Risk Modeling}{33}{3-4}{117--138}.
\newblock
\begin{APACrefDOI} \doi{10.1515/strm-2015-0005} \end{APACrefDOI}
\PrintBackRefs{\CurrentBib}

\bibitem [\protect \citeauthoryear {%
Battiston%
, Farmer%
\BCBL {}\ \protect \BOthers {.}}{%
Battiston%
, Farmer%
\BCBL {}\ \protect \BOthers {.}}{%
{\protect \APACyear {2016}}%
}]{%
battiston2016complexity}
\APACinsertmetastar {%
battiston2016complexity}%
\begin{APACrefauthors}%
Battiston, S.%
, Farmer, J\BPBI D.%
, Flache, A.%
, Garlaschelli, D.%
, Haldane, A\BPBI G.%
, Heesterbeek, H.%
\BDBL {}Scheffer, M.%
\end{APACrefauthors}%
\unskip\
\newblock
\APACrefYearMonthDay{2016}{}{}.
\newblock
{\BBOQ}\APACrefatitle {Complexity theory and financial regulation} {Complexity
  theory and financial regulation}.{\BBCQ}
\newblock
\APACjournalVolNumPages{Science}{351}{6275}{818--819}.
\newblock
\begin{APACrefDOI} \doi{10.1126/science.aad0299} \end{APACrefDOI}
\PrintBackRefs{\CurrentBib}

\bibitem [\protect \citeauthoryear {%
Battiston%
\ \BBA {} Mart\'nez-Jaramillo%
}{%
Battiston%
\ \BBA {} Mart\'nez-Jaramillo%
}{%
{\protect \APACyear {2018}}%
}]{%
battiston2018financial}
\APACinsertmetastar {%
battiston2018financial}%
\begin{APACrefauthors}%
Battiston, S.%
\BCBT {}\ \BBA {} Mart\'nez-Jaramillo, S.%
\end{APACrefauthors}%
\unskip\
\newblock
\APACrefYearMonthDay{2018}{}{}.
\newblock
{\BBOQ}\APACrefatitle {Financial networks and stress testing: Challenges and
  new research avenues for systemic risk analysis and financial stability
  implications} {Financial networks and stress testing: Challenges and new
  research avenues for systemic risk analysis and financial stability
  implications}.{\BBCQ}
\newblock
\APACjournalVolNumPages{Journal of Financial Stability}{35}{}{6--16}.
\newblock
\begin{APACrefDOI} \doi{10.1016/j.jfs.2018.03.010} \end{APACrefDOI}
\PrintBackRefs{\CurrentBib}

\bibitem [\protect \citeauthoryear {%
Battiston%
, Puliga%
, Kaushik%
, Tasca%
\BCBL {}\ \BBA {} Caldarelli%
}{%
Battiston%
\ \protect \BOthers {.}}{%
{\protect \APACyear {2012}}%
}]{%
battiston2012debtrank}
\APACinsertmetastar {%
battiston2012debtrank}%
\begin{APACrefauthors}%
Battiston, S.%
, Puliga, M.%
, Kaushik, R.%
, Tasca, P.%
\BCBL {}\ \BBA {} Caldarelli, G.%
\end{APACrefauthors}%
\unskip\
\newblock
\APACrefYearMonthDay{2012}{}{}.
\newblock
{\BBOQ}\APACrefatitle {Debtrank: Too central to fail? Financial networks, the
  FED and systemic risk} {Debtrank: Too central to fail? financial networks,
  the fed and systemic risk}.{\BBCQ}
\newblock
\APACjournalVolNumPages{Scientific Reports}{2}{}{541}.
\newblock
\begin{APACrefDOI} \doi{10.1038/srep00541} \end{APACrefDOI}
\PrintBackRefs{\CurrentBib}

\bibitem [\protect \citeauthoryear {%
Benoit%
, Colliard%
, Hurlin%
\BCBL {}\ \BBA {} P\'erignon%
}{%
Benoit%
\ \protect \BOthers {.}}{%
{\protect \APACyear {2017}}%
}]{%
benoit2017where}
\APACinsertmetastar {%
benoit2017where}%
\begin{APACrefauthors}%
Benoit, S.%
, Colliard, J\BHBI E.%
, Hurlin, C.%
\BCBL {}\ \BBA {} P\'erignon, C.%
\end{APACrefauthors}%
\unskip\
\newblock
\APACrefYearMonthDay{2017}{}{}.
\newblock
{\BBOQ}\APACrefatitle {Where the risks lie: A survey on systemic risk} {Where
  the risks lie: A survey on systemic risk}.{\BBCQ}
\newblock
\APACjournalVolNumPages{Review of Finance}{21}{1}{109--152}.
\newblock
\begin{APACrefDOI} \doi{10.1093/rof/rfw026} \end{APACrefDOI}
\PrintBackRefs{\CurrentBib}

\bibitem [\protect \citeauthoryear {%
Boss%
, Elsinger%
, Summer%
\BCBL {}\ \BBA {} Thurner%
}{%
Boss%
\ \protect \BOthers {.}}{%
{\protect \APACyear {2004}}%
}]{%
boss2004network}
\APACinsertmetastar {%
boss2004network}%
\begin{APACrefauthors}%
Boss, M.%
, Elsinger, H.%
, Summer, M.%
\BCBL {}\ \BBA {} Thurner, S.%
\end{APACrefauthors}%
\unskip\
\newblock
\APACrefYearMonthDay{2004}{}{}.
\newblock
{\BBOQ}\APACrefatitle {Network topology of the interbank market} {Network
  topology of the interbank market}.{\BBCQ}
\newblock
\APACjournalVolNumPages{Quantitative Finance}{4}{6}{677-684}.
\newblock
\begin{APACrefDOI} \doi{10.1080/14697680400020325} \end{APACrefDOI}
\PrintBackRefs{\CurrentBib}

\bibitem [\protect \citeauthoryear {%
Caccioli%
, Barucca%
\BCBL {}\ \BBA {} Kobayashi%
}{%
Caccioli%
\ \protect \BOthers {.}}{%
{\protect \APACyear {2018}}%
}]{%
caccioli2018network}
\APACinsertmetastar {%
caccioli2018network}%
\begin{APACrefauthors}%
Caccioli, F.%
, Barucca, P.%
\BCBL {}\ \BBA {} Kobayashi, T.%
\end{APACrefauthors}%
\unskip\
\newblock
\APACrefYearMonthDay{2018}{}{}.
\newblock
{\BBOQ}\APACrefatitle {Network models of financial systemic risk: A review}
  {Network models of financial systemic risk: A review}.{\BBCQ}
\newblock
\APACjournalVolNumPages{Journal of Computational Social
  Science}{1}{1}{81--114}.
\newblock
\begin{APACrefDOI} \doi{10.1007/s42001-017-0008-3} \end{APACrefDOI}
\PrintBackRefs{\CurrentBib}

\bibitem [\protect \citeauthoryear {%
Caccioli%
, Shrestha%
, Moore%
\BCBL {}\ \BBA {} Farmer%
}{%
Caccioli%
\ \protect \BOthers {.}}{%
{\protect \APACyear {2014}}%
}]{%
caccioli2014stability}
\APACinsertmetastar {%
caccioli2014stability}%
\begin{APACrefauthors}%
Caccioli, F.%
, Shrestha, M.%
, Moore, C.%
\BCBL {}\ \BBA {} Farmer, J\BPBI D.%
\end{APACrefauthors}%
\unskip\
\newblock
\APACrefYearMonthDay{2014}{}{}.
\newblock
{\BBOQ}\APACrefatitle {Stability analysis of financial contagion due to
  overlapping portfolios} {Stability analysis of financial contagion due to
  overlapping portfolios}.{\BBCQ}
\newblock
\APACjournalVolNumPages{Journal of Banking {\&} Finance}{46}{}{233--245}.
\newblock
\begin{APACrefDOI} \doi{10.1016/j.jbankfin.2014.05.021} \end{APACrefDOI}
\PrintBackRefs{\CurrentBib}

\bibitem [\protect \citeauthoryear {%
Cimini%
\ \BBA {} Serri%
}{%
Cimini%
\ \BBA {} Serri%
}{%
{\protect \APACyear {2016}}%
}]{%
cimini2016entangling}
\APACinsertmetastar {%
cimini2016entangling}%
\begin{APACrefauthors}%
Cimini, G.%
\BCBT {}\ \BBA {} Serri, M.%
\end{APACrefauthors}%
\unskip\
\newblock
\APACrefYearMonthDay{2016}{}{}.
\newblock
{\BBOQ}\APACrefatitle {Entangling credit and funding shocks in interbank
  markets} {Entangling credit and funding shocks in interbank markets}.{\BBCQ}
\newblock
\APACjournalVolNumPages{PLoS ONE}{11}{8}{e0161642}.
\newblock
\begin{APACrefDOI} \doi{10.1371/journal.pone.0161642} \end{APACrefDOI}
\PrintBackRefs{\CurrentBib}

\bibitem [\protect \citeauthoryear {%
Cimini%
, Squartini%
, Garlaschelli%
\BCBL {}\ \BBA {} Gabrielli%
}{%
Cimini%
\ \protect \BOthers {.}}{%
{\protect \APACyear {2015}}%
}]{%
cimini2015systemic}
\APACinsertmetastar {%
cimini2015systemic}%
\begin{APACrefauthors}%
Cimini, G.%
, Squartini, T.%
, Garlaschelli, D.%
\BCBL {}\ \BBA {} Gabrielli, A.%
\end{APACrefauthors}%
\unskip\
\newblock
\APACrefYearMonthDay{2015}{}{}.
\newblock
{\BBOQ}\APACrefatitle {Systemic risk analysis on reconstructed economic and
  financial networks} {Systemic risk analysis on reconstructed economic and
  financial networks}.{\BBCQ}
\newblock
\APACjournalVolNumPages{Scientific Reports}{5}{}{15758}.
\newblock
\begin{APACrefDOI} \doi{10.1038/srep15758} \end{APACrefDOI}
\PrintBackRefs{\CurrentBib}

\bibitem [\protect \citeauthoryear {%
Cimini%
\ \protect \BOthers {.}}{%
Cimini%
\ \protect \BOthers {.}}{%
{\protect \APACyear {2019}}%
}]{%
cimini2019statistical}
\APACinsertmetastar {%
cimini2019statistical}%
\begin{APACrefauthors}%
Cimini, G.%
, Squartini, T.%
, Saracco, F.%
, Garlaschelli, D.%
, Gabrielli, A.%
\BCBL {}\ \BBA {} Caldarelli, G.%
\end{APACrefauthors}%
\unskip\
\newblock
\APACrefYearMonthDay{2019}{}{}.
\newblock
{\BBOQ}\APACrefatitle {The statistical physics of real-world networks} {The
  statistical physics of real-world networks}.{\BBCQ}
\newblock
\APACjournalVolNumPages{Nature Reviews Physics}{1}{1}{58-71}.
\newblock
\begin{APACrefDOI} \doi{10.1038/s42254-018-0002-6} \end{APACrefDOI}
\PrintBackRefs{\CurrentBib}

\bibitem [\protect \citeauthoryear {%
Cocco%
, Gomes%
\BCBL {}\ \BBA {} Martins%
}{%
Cocco%
\ \protect \BOthers {.}}{%
{\protect \APACyear {2009}}%
}]{%
cocco2009lending}
\APACinsertmetastar {%
cocco2009lending}%
\begin{APACrefauthors}%
Cocco, J\BPBI F.%
, Gomes, F\BPBI J.%
\BCBL {}\ \BBA {} Martins, N\BPBI C.%
\end{APACrefauthors}%
\unskip\
\newblock
\APACrefYearMonthDay{2009}{}{}.
\newblock
{\BBOQ}\APACrefatitle {Lending relationships in the interbank market} {Lending
  relationships in the interbank market}.{\BBCQ}
\newblock
\APACjournalVolNumPages{Journal of Financial Intermediation}{18}{1}{24-48}.
\newblock
\begin{APACrefDOI} \doi{10.1016/j.jfi.2008.06.003} \end{APACrefDOI}
\PrintBackRefs{\CurrentBib}

\bibitem [\protect \citeauthoryear {%
Cont%
\ \BBA {} Wagalath%
}{%
Cont%
\ \BBA {} Wagalath%
}{%
{\protect \APACyear {2016}}%
}]{%
cont2016fire}
\APACinsertmetastar {%
cont2016fire}%
\begin{APACrefauthors}%
Cont, R.%
\BCBT {}\ \BBA {} Wagalath, L.%
\end{APACrefauthors}%
\unskip\
\newblock
\APACrefYearMonthDay{2016}{}{}.
\newblock
{\BBOQ}\APACrefatitle {Fire sales forensics: Measuring endogenous risk} {Fire
  sales forensics: Measuring endogenous risk}.{\BBCQ}
\newblock
\APACjournalVolNumPages{Mathematical Finance}{26}{4}{835--866}.
\newblock
\begin{APACrefDOI} \doi{10.1111/mafi.12071} \end{APACrefDOI}
\PrintBackRefs{\CurrentBib}

\bibitem [\protect \citeauthoryear {%
Covi%
, Montagna%
\BCBL {}\ \BBA {} Torri%
}{%
Covi%
\ \protect \BOthers {.}}{%
{\protect \APACyear {2019}}%
}]{%
covi2019economic}
\APACinsertmetastar {%
covi2019economic}%
\begin{APACrefauthors}%
Covi, G.%
, Montagna, M.%
\BCBL {}\ \BBA {} Torri, G.%
\end{APACrefauthors}%
\unskip\
\newblock
\APACrefYearMonthDay{2019}{May}{}.
\newblock
{\BBOQ}\APACrefatitle {Economic shocks and contagion in the euro area banking
  sector: A new micro-structural approach} {Economic shocks and contagion in
  the euro area banking sector: A new micro-structural approach}.{\BBCQ}
\newblock
\BIn{} \APACrefbtitle {Financial Stability Review.} {Financial stability
  review.}
\PrintBackRefs{\CurrentBib}

\bibitem [\protect \citeauthoryear {%
Eisenberg%
\ \BBA {} Noe%
}{%
Eisenberg%
\ \BBA {} Noe%
}{%
{\protect \APACyear {2001}}%
}]{%
eisenberg2001systemic}
\APACinsertmetastar {%
eisenberg2001systemic}%
\begin{APACrefauthors}%
Eisenberg, L.%
\BCBT {}\ \BBA {} Noe, T\BPBI H.%
\end{APACrefauthors}%
\unskip\
\newblock
\APACrefYearMonthDay{2001}{}{}.
\newblock
{\BBOQ}\APACrefatitle {Systemic risk in financial systems} {Systemic risk in
  financial systems}.{\BBCQ}
\newblock
\APACjournalVolNumPages{Management Science}{47}{2}{236--249}.
\newblock
\begin{APACrefDOI} \doi{10.2139/ssrn.173249} \end{APACrefDOI}
\PrintBackRefs{\CurrentBib}

\bibitem [\protect \citeauthoryear {%
Furfine%
}{%
Furfine%
}{%
{\protect \APACyear {2003}}%
}]{%
furfine2003interbank}
\APACinsertmetastar {%
furfine2003interbank}%
\begin{APACrefauthors}%
Furfine, C.%
\end{APACrefauthors}%
\unskip\
\newblock
\APACrefYearMonthDay{2003}{}{}.
\newblock
{\BBOQ}\APACrefatitle {Interbank exposures: Quantifying the risk of contagion}
  {Interbank exposures: Quantifying the risk of contagion}.{\BBCQ}
\newblock
\APACjournalVolNumPages{Journal of Money, Credit, and
  Banking}{35}{1}{111--128}.
\newblock
\begin{APACrefDOI} \doi{10.1353/mcb.2003.0004} \end{APACrefDOI}
\PrintBackRefs{\CurrentBib}

\bibitem [\protect \citeauthoryear {%
Gai%
\ \BBA {} Kapadia%
}{%
Gai%
\ \BBA {} Kapadia%
}{%
{\protect \APACyear {2010}}%
}]{%
gai2010contagion}
\APACinsertmetastar {%
gai2010contagion}%
\begin{APACrefauthors}%
Gai, P.%
\BCBT {}\ \BBA {} Kapadia, S.%
\end{APACrefauthors}%
\unskip\
\newblock
\APACrefYearMonthDay{2010}{}{}.
\newblock
{\BBOQ}\APACrefatitle {Contagion in financial networks} {Contagion in financial
  networks}.{\BBCQ}
\newblock
\APACjournalVolNumPages{Proceedings of the Royal Society of London A:
  Mathematical, Physical and Engineering Sciences}{466}{2120}{2401--2423}.
\newblock
\begin{APACrefDOI} \doi{10.1098/rspa.2009.0410} \end{APACrefDOI}
\PrintBackRefs{\CurrentBib}

\bibitem [\protect \citeauthoryear {%
Georg%
}{%
Georg%
}{%
{\protect \APACyear {2013}}%
}]{%
georg2013effect}
\APACinsertmetastar {%
georg2013effect}%
\begin{APACrefauthors}%
Georg, C\BHBI P.%
\end{APACrefauthors}%
\unskip\
\newblock
\APACrefYearMonthDay{2013}{}{}.
\newblock
{\BBOQ}\APACrefatitle {The effect of the interbank network structure on
  contagion and common shocks} {The effect of the interbank network structure
  on contagion and common shocks}.{\BBCQ}
\newblock
\APACjournalVolNumPages{Journal of Banking and Finance}{37}{7}{2216--2228}.
\newblock
\begin{APACrefDOI} \doi{10.1016/j.jbankfin.2013.02.032} \end{APACrefDOI}
\PrintBackRefs{\CurrentBib}

\bibitem [\protect \citeauthoryear {%
Greenwood%
, Landier%
\BCBL {}\ \BBA {} Thesmar%
}{%
Greenwood%
\ \protect \BOthers {.}}{%
{\protect \APACyear {2015}}%
}]{%
greenwood2015vulnerable}
\APACinsertmetastar {%
greenwood2015vulnerable}%
\begin{APACrefauthors}%
Greenwood, R.%
, Landier, A.%
\BCBL {}\ \BBA {} Thesmar, D.%
\end{APACrefauthors}%
\unskip\
\newblock
\APACrefYearMonthDay{2015}{}{}.
\newblock
{\BBOQ}\APACrefatitle {Vulnerable banks} {Vulnerable banks}.{\BBCQ}
\newblock
\APACjournalVolNumPages{Journal of Financial Economics}{115}{3}{471--485}.
\newblock
\begin{APACrefDOI} \doi{10.1016/j.jfineco.2014.11.006} \end{APACrefDOI}
\PrintBackRefs{\CurrentBib}

\bibitem [\protect \citeauthoryear {%
Gualdi%
, Cimini%
, Primicerio%
, Clemente%
\BCBL {}\ \BBA {} Challet%
}{%
Gualdi%
\ \protect \BOthers {.}}{%
{\protect \APACyear {2016}}%
}]{%
gualdi2016statistically}
\APACinsertmetastar {%
gualdi2016statistically}%
\begin{APACrefauthors}%
Gualdi, S.%
, Cimini, G.%
, Primicerio, K.%
, Clemente, R\BPBI D.%
\BCBL {}\ \BBA {} Challet, D.%
\end{APACrefauthors}%
\unskip\
\newblock
\APACrefYearMonthDay{2016}{}{}.
\newblock
{\BBOQ}\APACrefatitle {Statistically validated network of portfolio overlaps
  and systemic risk} {Statistically validated network of portfolio overlaps and
  systemic risk}.{\BBCQ}
\newblock
\APACjournalVolNumPages{Scientific Reports}{6}{}{39467}.
\newblock
\begin{APACrefDOI} \doi{10.1038/srep39467} \end{APACrefDOI}
\PrintBackRefs{\CurrentBib}

\bibitem [\protect \citeauthoryear {%
Haldane%
\ \BBA {} May%
}{%
Haldane%
\ \BBA {} May%
}{%
{\protect \APACyear {2011}}%
}]{%
haldane2011systemic}
\APACinsertmetastar {%
haldane2011systemic}%
\begin{APACrefauthors}%
Haldane, A\BPBI G.%
\BCBT {}\ \BBA {} May, R\BPBI M.%
\end{APACrefauthors}%
\unskip\
\newblock
\APACrefYearMonthDay{2011}{}{}.
\newblock
{\BBOQ}\APACrefatitle {Systemic risk in banking ecosystems} {Systemic risk in
  banking ecosystems}.{\BBCQ}
\newblock
\APACjournalVolNumPages{Nature}{469}{}{351--355}.
\newblock
\begin{APACrefDOI} \doi{10.1038/nature09659} \end{APACrefDOI}
\PrintBackRefs{\CurrentBib}

\bibitem [\protect \citeauthoryear {%
Hurd%
, Gleeson%
\BCBL {}\ \BBA {} Melnik%
}{%
Hurd%
\ \protect \BOthers {.}}{%
{\protect \APACyear {2017}}%
}]{%
hurd2017framework}
\APACinsertmetastar {%
hurd2017framework}%
\begin{APACrefauthors}%
Hurd, T\BPBI R.%
, Gleeson, J\BPBI P.%
\BCBL {}\ \BBA {} Melnik, S.%
\end{APACrefauthors}%
\unskip\
\newblock
\APACrefYearMonthDay{2017}{}{}.
\newblock
{\BBOQ}\APACrefatitle {A framework for analyzing contagion in assortative
  banking networks} {A framework for analyzing contagion in assortative banking
  networks}.{\BBCQ}
\newblock
\APACjournalVolNumPages{PLoS ONE}{12}{2}{e0170579}.
\newblock
\begin{APACrefDOI} \doi{10.1371/journal.pone.0170579} \end{APACrefDOI}
\PrintBackRefs{\CurrentBib}

\bibitem [\protect \citeauthoryear {%
H{\"u}ser%
}{%
H{\"u}ser%
}{%
{\protect \APACyear {2015}}%
}]{%
huser2015too}
\APACinsertmetastar {%
huser2015too}%
\begin{APACrefauthors}%
H{\"u}ser, A\BHBI C.%
\end{APACrefauthors}%
\unskip\
\newblock
\APACrefYearMonthDay{2015}{}{}.
\newblock
{\BBOQ}\APACrefatitle {Too interconnected to fail: A survey of the interbank
  networks literature} {Too interconnected to fail: A survey of the interbank
  networks literature}.{\BBCQ}
\newblock
\APACjournalVolNumPages{Journal of Network Theory In Finance}{1}{3}{1--50}.
\newblock
\begin{APACrefDOI} \doi{10.21314/JNTF.2015.001} \end{APACrefDOI}
\PrintBackRefs{\CurrentBib}

\bibitem [\protect \citeauthoryear {%
Krause%
, \v{S}tefan\v{c}i\'c%
, Zlati\'c%
\BCBL {}\ \BBA {} Caldarelli%
}{%
Krause%
\ \protect \BOthers {.}}{%
{\protect \APACyear {2019}}%
}]{%
krause2019controlling}
\APACinsertmetastar {%
krause2019controlling}%
\begin{APACrefauthors}%
Krause, S\BPBI M.%
, \v{S}tefan\v{c}i\'c, H.%
, Zlati\'c, V.%
\BCBL {}\ \BBA {} Caldarelli, G.%
\end{APACrefauthors}%
\unskip\
\newblock
\APACrefYearMonthDay{2019}{}{}.
\newblock
\APACrefbtitle {Controlling systemic risk -- network structures that minimize
  it and node properties to calculate it.} {Controlling systemic risk --
  network structures that minimize it and node properties to calculate it.}
\newblock
\APAChowpublished {\url{https://arxiv.org/abs/1902.08483}}.
\PrintBackRefs{\CurrentBib}

\bibitem [\protect \citeauthoryear {%
Le{\'o}n%
\ \BBA {} Berndsen%
}{%
Le{\'o}n%
\ \BBA {} Berndsen%
}{%
{\protect \APACyear {2014}}%
}]{%
leon2014rethinking}
\APACinsertmetastar {%
leon2014rethinking}%
\begin{APACrefauthors}%
Le{\'o}n, C.%
\BCBT {}\ \BBA {} Berndsen, R\BPBI J.%
\end{APACrefauthors}%
\unskip\
\newblock
\APACrefYearMonthDay{2014}{}{}.
\newblock
{\BBOQ}\APACrefatitle {Rethinking financial stability: Challenges arising from
  financial networks' modular scale-free architecture} {Rethinking financial
  stability: Challenges arising from financial networks' modular scale-free
  architecture}.{\BBCQ}
\newblock
\APACjournalVolNumPages{Journal of Financial Stability}{15}{}{241-256}.
\newblock
\begin{APACrefDOI} \doi{10.1016/j.jfs.2014.10.006} \end{APACrefDOI}
\PrintBackRefs{\CurrentBib}

\bibitem [\protect \citeauthoryear {%
Mastromatteo%
, Zarinelli%
\BCBL {}\ \BBA {} Marsili%
}{%
Mastromatteo%
\ \protect \BOthers {.}}{%
{\protect \APACyear {2012}}%
}]{%
mastromatteo2012reconstruction}
\APACinsertmetastar {%
mastromatteo2012reconstruction}%
\begin{APACrefauthors}%
Mastromatteo, I.%
, Zarinelli, E.%
\BCBL {}\ \BBA {} Marsili, M.%
\end{APACrefauthors}%
\unskip\
\newblock
\APACrefYearMonthDay{2012}{}{}.
\newblock
{\BBOQ}\APACrefatitle {Reconstruction of financial networks for robust
  estimation of systemic risk} {Reconstruction of financial networks for robust
  estimation of systemic risk}.{\BBCQ}
\newblock
\APACjournalVolNumPages{Journal of Statistical Mechanics: Theory and
  Experiment}{2012}{03}{P03011}.
\newblock
\begin{APACrefDOI} \doi{10.1088/1742-5468/2012/03/P03011} \end{APACrefDOI}
\PrintBackRefs{\CurrentBib}

\bibitem [\protect \citeauthoryear {%
Montagna%
\ \BBA {} Lux%
}{%
Montagna%
\ \BBA {} Lux%
}{%
{\protect \APACyear {2017}}%
}]{%
montagna2017contagion}
\APACinsertmetastar {%
montagna2017contagion}%
\begin{APACrefauthors}%
Montagna, M.%
\BCBT {}\ \BBA {} Lux, T.%
\end{APACrefauthors}%
\unskip\
\newblock
\APACrefYearMonthDay{2017}{}{}.
\newblock
{\BBOQ}\APACrefatitle {Contagion risk in the interbank market: A probabilistic
  approach to cope with incomplete structural information} {Contagion risk in
  the interbank market: A probabilistic approach to cope with incomplete
  structural information}.{\BBCQ}
\newblock
\APACjournalVolNumPages{Quantitative Finance}{17}{1}{101--120}.
\newblock
\begin{APACrefDOI} \doi{10.1080/14697688.2016.1178855} \end{APACrefDOI}
\PrintBackRefs{\CurrentBib}

\bibitem [\protect \citeauthoryear {%
Nier%
, Yang%
, Yorulmazer%
\BCBL {}\ \BBA {} Alentorn%
}{%
Nier%
\ \protect \BOthers {.}}{%
{\protect \APACyear {2007}}%
}]{%
nier2007network}
\APACinsertmetastar {%
nier2007network}%
\begin{APACrefauthors}%
Nier, E\BPBI W.%
, Yang, J.%
, Yorulmazer, T.%
\BCBL {}\ \BBA {} Alentorn, A.%
\end{APACrefauthors}%
\unskip\
\newblock
\APACrefYearMonthDay{2007}{}{}.
\newblock
{\BBOQ}\APACrefatitle {Network models and financial stability} {Network models
  and financial stability}.{\BBCQ}
\newblock
\APACjournalVolNumPages{Journal of Economic Dynamics and
  Control}{31}{6}{2033-2060}.
\newblock
\begin{APACrefDOI} \doi{10.1016/j.jedc.2007.01.014} \end{APACrefDOI}
\PrintBackRefs{\CurrentBib}

\bibitem [\protect \citeauthoryear {%
Pichler%
, Poledna%
\BCBL {}\ \BBA {} Thurner%
}{%
Pichler%
\ \protect \BOthers {.}}{%
{\protect \APACyear {2018}}%
}]{%
pichler2018systemic}
\APACinsertmetastar {%
pichler2018systemic}%
\begin{APACrefauthors}%
Pichler, A.%
, Poledna, S.%
\BCBL {}\ \BBA {} Thurner, S.%
\end{APACrefauthors}%
\unskip\
\newblock
\APACrefYearMonthDay{2018}{}{}.
\newblock
\APACrefbtitle {Systemic-risk-efficient asset allocation: Minimization of
  systemic risk as a network optimization problem.} {Systemic-risk-efficient
  asset allocation: Minimization of systemic risk as a network optimization
  problem.}
\newblock
\APAChowpublished {\url{https://arxiv.org/abs/1801.10515}}.
\PrintBackRefs{\CurrentBib}

\bibitem [\protect \citeauthoryear {%
Poledna%
, Molina-Borboa%
, Mart\'inez-Jaramillo%
, van~der Leij%
\BCBL {}\ \BBA {} Thurner%
}{%
Poledna%
\ \protect \BOthers {.}}{%
{\protect \APACyear {2015}}%
}]{%
podelna2015multilayer}
\APACinsertmetastar {%
podelna2015multilayer}%
\begin{APACrefauthors}%
Poledna, S.%
, Molina-Borboa, J\BPBI L.%
, Mart\'inez-Jaramillo, S.%
, van~der Leij, M.%
\BCBL {}\ \BBA {} Thurner, S.%
\end{APACrefauthors}%
\unskip\
\newblock
\APACrefYearMonthDay{2015}{}{}.
\newblock
{\BBOQ}\APACrefatitle {The multi-layer network nature of systemic risk and its
  implications for the costs of financial crises} {The multi-layer network
  nature of systemic risk and its implications for the costs of financial
  crises}.{\BBCQ}
\newblock
\APACjournalVolNumPages{Journal of Financial Stability}{20}{}{70--81}.
\newblock
\begin{APACrefDOI} \doi{10.1016/j.jfs.2015.08.001} \end{APACrefDOI}
\PrintBackRefs{\CurrentBib}

\bibitem [\protect \citeauthoryear {%
Roukny%
, Bersini%
, Pirotte%
, Caldarelli%
\BCBL {}\ \BBA {} Battiston%
}{%
Roukny%
\ \protect \BOthers {.}}{%
{\protect \APACyear {2013}}%
}]{%
roukny2013default}
\APACinsertmetastar {%
roukny2013default}%
\begin{APACrefauthors}%
Roukny, T.%
, Bersini, H.%
, Pirotte, H.%
, Caldarelli, G.%
\BCBL {}\ \BBA {} Battiston, S.%
\end{APACrefauthors}%
\unskip\
\newblock
\APACrefYearMonthDay{2013}{}{}.
\newblock
{\BBOQ}\APACrefatitle {Default cascades in complex networks: Topology and
  systemic risk} {Default cascades in complex networks: Topology and systemic
  risk}.{\BBCQ}
\newblock
\APACjournalVolNumPages{Scientific Reports}{3}{}{2759}.
\newblock
\begin{APACrefDOI} \doi{10.1038/srep02759} \end{APACrefDOI}
\PrintBackRefs{\CurrentBib}

\bibitem [\protect \citeauthoryear {%
Squartini%
, Caldarelli%
, Cimini%
, Gabrielli%
\BCBL {}\ \BBA {} Garlaschelli%
}{%
Squartini%
\ \protect \BOthers {.}}{%
{\protect \APACyear {2018}}%
}]{%
squartini2018reconstruction}
\APACinsertmetastar {%
squartini2018reconstruction}%
\begin{APACrefauthors}%
Squartini, T.%
, Caldarelli, G.%
, Cimini, G.%
, Gabrielli, A.%
\BCBL {}\ \BBA {} Garlaschelli, D.%
\end{APACrefauthors}%
\unskip\
\newblock
\APACrefYearMonthDay{2018}{}{}.
\newblock
{\BBOQ}\APACrefatitle {Reconstruction methods for networks: The case of
  economic and financial systems} {Reconstruction methods for networks: The
  case of economic and financial systems}.{\BBCQ}
\newblock
\APACjournalVolNumPages{Physics Reports}{757}{}{1-47}.
\newblock
\begin{APACrefDOI} \doi{10.1016/j.physrep.2018.06.008} \end{APACrefDOI}
\PrintBackRefs{\CurrentBib}

\bibitem [\protect \citeauthoryear {%
Squartini%
, Cimini%
, Gabrielli%
\BCBL {}\ \BBA {} Garlaschelli%
}{%
Squartini%
\ \protect \BOthers {.}}{%
{\protect \APACyear {2017}}%
}]{%
squartini2017network}
\APACinsertmetastar {%
squartini2017network}%
\begin{APACrefauthors}%
Squartini, T.%
, Cimini, G.%
, Gabrielli, A.%
\BCBL {}\ \BBA {} Garlaschelli, D.%
\end{APACrefauthors}%
\unskip\
\newblock
\APACrefYearMonthDay{2017}{}{}.
\newblock
{\BBOQ}\APACrefatitle {Network reconstruction via density sampling} {Network
  reconstruction via density sampling}.{\BBCQ}
\newblock
\APACjournalVolNumPages{Applied Network Science}{2}{1}{3}.
\newblock
\begin{APACrefDOI} \doi{10.1007/s41109-017-0021-8} \end{APACrefDOI}
\PrintBackRefs{\CurrentBib}

\end{thebibliography}

\end{document}